\documentclass[fleqn,10pt]{wlscirep}

\usepackage[utf8]{inputenc}
\usepackage[T1]{fontenc}

\usepackage{amsmath}
\usepackage{amsfonts}
\usepackage{bbm}
\usepackage{bm}
\usepackage{lineno} 
\usepackage{euscript}
\renewcommand{\mathcal}{\EuScript}

\title{Dissimilarity measures for generalized Lotka-Volterra systems on networks}

\author[1]{Nicolás A. Márquez}
\author[2]{Maryam Chaib~De~Mares}
\author[1,*]{Alejandro P.~Riascos}
\affil[1]{Departamento de Física, Facultad de Ciencias, Universidad Nacional de Colombia, Bogotá 111321, Colombia}
\affil[2]{Departamento de Biología, Facultad de Ciencias, Universidad Nacional de Colombia, Bogotá 111321, Colombia}

\affil[*]{{\it Corresponding author email:} alperezri@unal.edu.co}


\begin{abstract}
In this paper, we introduce a general framework to quantify dissimilarities between generalized Lotka–Volterra  dynamical processes, ranging from classical predator–prey systems to multispecies communities interacting on networks. The proposed measures capture both transient and stationary dynamics, allowing systematic comparisons across systems with varying interaction parameters, network weights, or topologies. Our analysis shows that even subtle structural changes can lead to markedly distinct outcomes: in two-species systems, interaction strength and initial conditions strongly affect divergence, while in small directed networks, differences that are invisible at the adjacency-matrix level produce divergent dynamics. In modular networks, the fraction and distribution of negative interactions control the transition from stable to unstable dynamics, with localized perturbations within cliques yielding different global outcomes than distributed ones. Beyond structural variations, the framework also applies when modified processes follow distinct nonlinear equations, demonstrating its versatility. Taken together, these results highlight that dynamical dissimilarity measures provide a powerful tool to analyze robustness, detect structural sensitivity, and predict instabilities in nonlinear systems. More broadly, this approach supports the comparative analysis of biological systems, where complex interaction networks and nonlinear dynamics are central to stability and resilience.
\end{abstract}
\begin{document}

\flushbottom
\maketitle
%
%
\thispagestyle{empty}

\section{Introduction}
Ecosystems can be regarded as prototypical complex dynamical systems, where heterogeneous components interact across multiple spatial and temporal scales. This intricate interplay generates emergent properties and nonlinear dynamics that shape the functioning of the biosphere \cite{Riva2023EcologicalComplexity,SoleLevin2022EcologicalComplexity}. Network science provides a natural framework to represent species and habitats as interconnected systems, allowing the study of how structural features such as connectivity, modularity, and multilayer organization constrain ecological dynamics, influence the propagation of perturbations, and determine the stability of ecological processes \cite{Fortin2021NetworkEcology,Dormann2017PatternsEcologicalNetworks,Pilosof2017MultilayerEcologicalNetworks}.  
\\[2mm]
Within this perspective, the generalized Lotka–Volterra (gLV) equations extend the classical predator–prey formalism to multispecies communities, providing a flexible framework to describe coexistence, competition, and nonlinear feedbacks. They capture the essential mechanisms by which interaction networks generate collective behaviors in ecological systems. Recent studies have demonstrated how network topology, environmental stochasticity, and asymmetry shape persistence, diversity, and resilience \cite{PhysRevE.111.034408,PhysRevE.111.014318,Lastad2022}. These results establish gLV systems as universal models that bridge theoretical ecology and statistical physics, while computational approaches such as Bayesian inference and ensemble modeling further enhance their predictive capacity for ecosystem responses under environmental disturbances \cite{10.1111/2041-210X.70032}.  
\\[2mm]
The gLV framework has also become central in microbiome research, where it provides mechanistic insights into microbial interactions, stability, and resilience at the community level \cite{Stein2013,Bucci2016,Venturelli2018}. In this setting, each microbial taxon is represented as a dynamical variable whose growth depends on both intrinsic fitness and pairwise couplings. Applications range from synthetic microbial consortia in food fermentation \cite{10.1016/j.fm.2025.104767} to biomedical interventions targeting the gut microbiota \cite{Melvan_2025}, where dynamics and functional outcomes are strongly shaped by the underlying interaction matrix. By combining empirical data with Bayesian methods, recent work has demonstrated that gLV models can quantify uncertainty and generate robust predictions of microbial dynamics under external perturbations \cite{Gibson2025,castellanos2023}.  
\\[2mm]
Beyond ecology and microbiology, the gLV formalism has recently attracted attention as an analytical tool in neuroscience, offering a bridge between ecological dynamics and collective neural activity \cite{Lagzi2015,MeyerOrtmanns2023,Rabinovich2024}. Theoretical studies of interaction networks have shown that asymmetric connectivity and heterogeneous couplings naturally lead to rich dynamical regimes, including synchronization, chaos, metastability and antifragility \cite{10_1098_rspa_2024_0290,10_1016_j_physrep_2024_08_001,PhysRevE.110.044309,PoloGonzalez2025}. These findings resonate with principles governing brain networks and highlight the versatility of the gLV framework as a unifying model of collective dynamics across complex systems. In this broader context, gLV equations serve not only as a minimal model of interacting populations but also as a conceptual bridge linking ecology, microbiology, and neuroscience.
\\[2mm]
The wide applicability of gLV equations reveals the need for measures to compare systems that differ either in their interaction networks or in the differential equations governing their dynamics. To date, no such measure has been proposed. Most efforts in network science have focused on characterizing network topology, introducing numerous structural metrics to describe and compare networks across applications \cite{VespiBook,Barabasi_2016,Newman_Book2018}. These include classical distances such as Hamming \cite{Hamming1950} and Jaccard \cite{Jaccard1901,LEVANDOWSKY1971}, as well as more recent information–theoretic measures \cite{Bagrow2019}. Complementary spectral approaches evaluate smooth variations in global structure by tracking functions of the eigenvalues of the adjacency or Laplacian matrices \cite{Donnat_2018}. Other perspectives exploit graph kernels \cite{Jurman2015,Hammond2013,Scott2021}, or study dynamical features such as information flow \cite{DomenicoPRE2020}, transport \cite{RiascosDiffusion2023}, and synchronization \cite{RiascosKuramoto2023}. Despite these advances, a measure for comparing nonlinear dynamics, particularly population dynamics on networks, remains lacking, and would be central to understanding how network structure constrains dynamics.
\\[2mm]
In this contribution, we introduce a framework to quantify dissimilarities between gLV systems that differ in interaction parameters, network structure, or in the functional form of the governing equations. While existing approaches often rely on comparisons of interaction matrices, such measures do not necessarily capture how nonlinear dynamics translate these structural differences into distinct population trajectories. To address this limitation, our framework compares the resulting dynamics directly. Beginning with the classical predator--prey model, we extend the approach to multispecies networks with heterogeneous interactions and to cases with modified nonlinearities. This formulation enables systematic comparisons of both transient and asymptotic dynamics across a wide range of scenarios. Although the approach requires numerical integration of the governing equations, it provides information that cannot be inferred solely from structural metrics. Beyond its theoretical relevance, the methodology offers a tool to investigate robustness, detect structural sensitivity, and evaluate stability in complex dynamical systems, with potential applications in ecology, microbiology, and neuroscience. More broadly, the framework highlights the importance of incorporating dynamical information when comparing interacting complex systems.
\section{General Theory}
Motivated by the need to compare dynamical processes that differ in their interaction parameters, network structure, or governing equations, we now introduce the theoretical framework used throughout this work. We begin with the classical Lotka--Volterra predator--prey model, which provides the simplest setting to illustrate how differences in interaction parameters lead to divergent dynamics. Based on this formulation, we define a measure that quantifies the dissimilarity between two systems evolving from identical initial conditions. The approach is then generalized to multispecies systems interacting through networks, where the interaction matrix determines the structure and strength of interspecies couplings.

\subsection{The Lotka-Volterra model for predator prey systems}
\label{Sec_LVmodel}
The core of predicting and understanding ecological communities lies in quantifying the strength, sign, and origin of interspecies interactions, as well as their dependence on environmental context \cite{Muscarella2020}. In general, such interactions form an entire network of species, commonly referred to as a trophic network, which results in structurally complex communities that are challenging to analyze. Therefore, systems involving only two interacting species are typically considered as a starting point, to be later generalized to systems with $N$ species.
\\[2mm]
There are three main types of models used to describe systems involving two interacting species. The first is the predator--prey type, in which the growth rate of the prey population decreases, while that of the predator increases. The second type involves competition, where the growth rates of both populations are reduced. The third case corresponds to mutualism or symbiosis, where both populations experience increased growth rates \cite{Murray2002}. The first mathematical model proposed to describe such interactions was introduced by Vito Volterra in 1926 to describe the predation of one species upon another, aiming to explain the oscillatory behavior observed in certain fish populations  \cite{VOLTERRA1926}. Independently, Alfred J. Lotka proposed an equivalent model \cite{lotka1925elements}. In the Lotka--Volterra model, $X(t)$ denotes the prey population at time $t$, while $Y(t)$ refers to the predator population. The evolution of the system is given by \cite{Murray2002}
\begin{equation}
	\begin{aligned}
		\label{eq:LV}
		\dfrac {d X}{dt} &= \,X(t) \,\left(a-b\,Y(t)\right) ,\\[2mm] 
		\dfrac{d Y}{d t} &= \,Y(t)\,\left(c\,X(t)-d\right)
	\end{aligned}
\end{equation}
where \textit{a}, \textit{b}, \textit{c}, and \textit{d} are positive constants.
\\[2mm]
The model assumes that, in the absence of predators, the prey population exhibits unbounded growth following Malthusian dynamics \cite{Kot_2001}. Predation reduces the per capita growth rate of the prey, with the reduction proportional to the size of the interacting populations. In the absence of prey, the predator population declines exponentially. The positive contribution of the prey to the predator population is represented by a term describing the transfer of energy between species.
\\[2mm]
When non-dimensional variables are introduced via the substitutions
\begin{equation}
	\label{eq:LVA}
	u(\tau) = \dfrac{c\,X(t)}{d} \,,\qquad v(\tau) = \dfrac{b\, Y(t)}{a} \,,\qquad \tau= a \, t \,,\qquad\alpha =\dfrac{d}{a}\,,
\end{equation}
the dynamics described by Eqs.~(\ref{eq:LV}) result in the following dimensionless system
\begin{equation}
	\label{eq:LVA2}
	\begin{aligned}
		\dfrac{d\,u}{d\, \tau} &= u(\tau)\,\left(1-v(\tau)\right),\\
		\dfrac{d\, v}{d\,\tau} &= \, \alpha \, v(\tau)\,\left(u(\tau)-1\right).
	\end{aligned}   
\end{equation}
The system in Eq.~(\ref{eq:LVA2}) is a dimensionless equivalent of the Lotka--Volterra model in which a single parameter, $\alpha$, governs the predator--prey interaction strength and determines the balance between their growth and decline. This reduced form facilitates direct analysis of the system’s response to changes in $\alpha$.
\\[2mm]
An additional insight into the role of the parameter $\alpha$ in the dynamics of the system in Eq.~(\ref{eq:LVA2}) can be obtained by analyzing the linear stability of the equilibrium point. The system admits a coexistence equilibrium at $(u,v)=(1,1)$, corresponding to a state in which both populations remain constant in time. To study the behavior of small perturbations around this equilibrium, we consider the Jacobian matrix obtained from the linearization of Eq.~(\ref{eq:LVA2}),
\begin{equation}
	J(u,v)=
	\begin{pmatrix}
		\dfrac{\partial}{\partial u}\left[u(1-v)\right] &
		\dfrac{\partial}{\partial v}\left[u(1-v)\right] \\
		\dfrac{\partial}{\partial u}\left[\alpha v(u-1)\right] &
		\dfrac{\partial}{\partial v}\left[\alpha v(u-1)\right]
	\end{pmatrix}.
\end{equation}
Evaluating the derivatives yields
\begin{equation}
	J(u,v)=
	\begin{pmatrix}
		1-v & -u \\
		\alpha v & \alpha (u-1)
	\end{pmatrix}.
\end{equation}
At the coexistence equilibrium $(u,v)=(1,1)$ the Jacobian becomes
\begin{equation}
	J(1,1)=
	\begin{pmatrix}
		0 & -1 \\
		\alpha & 0
	\end{pmatrix}.
\end{equation}
The eigenvalues $\lambda$ of this matrix satisfy \cite{Murray2002}
\begin{equation}
	\lambda^2 + \alpha = 0,
\end{equation}
which gives $\lambda_{\pm} = \pm i\sqrt{\alpha}$. These purely imaginary eigenvalues indicate that the equilibrium corresponds to a neutrally stable center, consistent with the oscillatory behavior characteristic of the classical Lotka--Volterra system. The magnitude of the imaginary part determines the oscillation frequency of the populations, which scales as $\omega=\sqrt{\alpha}$. Therefore, larger values of $\alpha$ lead to faster oscillatory dynamics \cite{Murray2002}. 
\subsection{Dissimilarity in Lotka-Volterra type systems with two species }
\label{Sec_diss_LV2}
In this section, we introduce a measure of dissimilarity for comparing two Lotka--Volterra systems with identical species composition but different interaction parameters. Specifically, we consider a reference system $(u, v)$ and a modified system $(u^\star, v^\star)$, both governed by Eq.~(\ref{eq:LVA2}) with parameters $\alpha$ and $\alpha^\star$, respectively. The systems share the same initial conditions at $\tau = 0$: $u(0) = u^\star(0) = u_0$ and $v(0) = v^\star(0) = v_0$. The dissimilarity between them is quantified by $\mathcal{D}^{(\mathrm{LV})}(\tau)$, defined as
\begin{equation}
	\mathcal{D}^{(\mathrm{LV})}(\tau) \equiv \frac1{u_0 +v_0} \Bigg[ \Big|u(\tau)-u^{\star}(\tau)\Big| + \Big|v(\tau)-v^{\star}(\tau)\Big| \Bigg].
	\label{eq:D(t)}
\end{equation}    
This quantity captures the evolution of the population differences between the two systems over time. In particular, $\mathcal{D}^{(\mathrm{LV})}(0) = 0$ since both systems share the same initial conditions. It is also useful to track the maximum difference observed up to time $\tau$, denoted $\mathcal{D}^{(\mathrm{LV})}_{\mathrm{max}}(\tau)$, and defined as
\begin{equation}
	\mathcal{D}^{(\mathrm{LV})}_{\mathrm{max}}(\tau)\equiv \mathrm{max}
	\left(\{\mathcal{D}^{(\mathrm{LV})}(t^\prime): 0\leq t^\prime\leq \tau\}\right).
	\label{eq:dmax}
\end{equation}
By employing $\mathcal{D}^{(\mathrm{LV})}(\tau)$ and $\mathcal{D}^{(\mathrm{LV})}_{\mathrm{max}}(\tau)$, our objective is to characterize the temporal differences between the population dynamics, with particular focus on the effects of varying the governing parameter $\alpha$.
\subsection{Lotka--Volterra type processes on networks}
In the mid-1950s, ecologists such as Odum and MacArthur proposed that ecosystems with greater species richness tend to exhibit higher stability than those with lower biodiversity \cite{Liu2023}. This idea was formalized in 1972 by May, who employed random matrix theory to analyze the stability of systems of interacting species \cite{may1972}. The framework considers a system of $N$ species $x_i$, with $i=1,2,\ldots,N$, interacting until they reach an equilibrium state $\bm{x}^* = \left( x_{1}^{*}, \ldots, x_{N}^{*} \right)$, such that
\begin{equation}
	\dfrac{d x_i}{d t} = x_i \Phi_{i}(\textbf{x}^*) = 0,
	\label{eq:Akjouj}
\end{equation}
where $\Phi_i(\bm{x})$ is assumed to be linear, with domain $(\mathbb{R}_{>0})^N$. In addition, it is assumed that in the vicinity of the equilibrium point, the Jacobian matrix $\mathcal{J}(\bm{x}, \Phi(\bm{x}))$ takes the form
\begin{align*}
	\mathcal{J}_{ii} & =  {- \mathbbm{I}+\mathcal{M}_{ii} C_{ii}} \\  \mathcal{J}_{ij} & = \mathcal{M}_{ij} C_{ij},
\end{align*}   
where $\mathbbm{I}$ is the identity matrix of size $N \times N$.
\\[2mm]
Furthermore, $\mathcal{M}_{ij}$ is a random variable with variance \textit{V}, while $C_{ij}$ follows a Bernoulli distribution with parameter \textit{C}. The main conclusion of the study is that the equilibrium is stable when $NCV < 1$, and unstable when $NCV > 1$ \cite{may1972,Akjouj2024}.
\\[2mm]
This stability criterion was extended by Allesina and Tang \cite{Allesina2012}, who considered matrix elements $(a_{ij}, a_{ji})$ drawn from a joint normal distribution, $(a_{ij}, a_{ji}) \sim \mathcal{N}(\bm{0}, \bm{\Sigma})$.  
Here, $\mathcal{N}(\mu, \sigma)$ denotes a univariate normal distribution with mean $\mu$ and variance $\sigma^2$, while $\mathcal{N}(\bm{0}, \bm{\Sigma})$ refers to a bivariate normal distribution with zero mean and covariance matrix $\bm{\Sigma}$, which encodes the correlation between $a_{ij}$ and $a_{ji}$ through the matrix \cite{Liu2023}
\begin{equation*}
	\bm{\Sigma}=
	\left[
	\begin{array}{cc}
		1 & \rho\\
		\rho & 1
	\end{array}
	\right].
\end{equation*}
Here, $\rho < 1$ reflects a predominance of predator--prey interactions, whereas $\rho > 1$ indicates a higher prevalence of mutualistic or competitive interactions. This generalization yields the stability criterion $NCV(1+\rho)^2 < 1$, implying that predator--prey dominance enhances the \emph{probability of stability} while an excess of mutualistic or competitive interactions reduces it \cite{Liu2023}. In complex ecosystems, characterized by many interacting components and continuous exchange with the environment, the governing equations must balance the inclusion of essential system features with the simplicity required to reproduce observed dynamics, with emphasis on a comprehensive understanding of the underlying processes.
\\[2mm]
In the function $\Phi_i(\bm{x}) = r_i - \sum_{j=1}^{N} \Lambda_{ij} x_j(t)$, defined in Eq.~(\ref{eq:Akjouj}), the sign of $\Lambda_{ij}$ determines the nature of the interaction: $\Lambda_{ij} > 0$ denotes interspecific competition, while $\Lambda_{ij} < 0$ corresponds to mutualism. Real ecosystems, however, rarely exhibit purely competitive or purely mutualistic relationships; mixtures of interaction types are common, including predation, amensalism, and commensalism. In general, these interactions are asymmetric, i.e., $\Lambda_{ij} \neq \Lambda_{ji}$.
\\[2mm]
The dynamics associated with ecological networks can be studied numerically using a gLV model, in which the population dynamics of each species is described by
\begin{equation}
	\label{eq:GLV_maryam}
	\frac{d {x}_i(t)}{dt} = x_i(t) \left(r_i -\sum_{j=1}^{N} \Lambda_{ij}\,x_j(t) \right).
\end{equation}
Here, $N$ denotes the number of species, and the elements $\Lambda_{ij}$ of the matrix $\bm{\Lambda}$ represent the interaction between species $i$ and species $j$. It is assumed that $r_i = r$, i.e., all species share the same intrinsic growth rate. This simplification is made in order to isolate the effects of interspecies interactions.
\\[2mm]
Additionally, we introduce the rescaled time variable $\tau = r t$, which implies $d\tau = r\,dt$. In this way,
\begin{equation}
	\dfrac{dx_i}{d\tau} = \dfrac{dx_i}{dt} \left( \dfrac{dt}{d\tau} \right) \,.
\end{equation}
Substituting into Eq.~(\ref{eq:GLV_maryam}) yields
\begin{equation}
	\dfrac{dx_i}{d\tau} = \left( \dfrac{1}{r} \right) \left[ x_i \left( r - \sum_{j=1}^{N} \Lambda_{ij} x_j \right) \right] \,,
\end{equation}
which simplifies to
\begin{equation}
	\dfrac{dx_i}{d\tau} = x_i \left( 1 - \sum_{j=1}^{N} \dfrac{\Lambda_{ij} x_j}{r} \right) \,.
\end{equation}
Furthermore, under the rescaling $x_i = r \mathcal{N}_i$, we have
\begin{equation}
	\dfrac{d (r \mathcal{N}_i)}{d\tau} = r \mathcal{N}_i \left( 1 - \sum_{j=1}^{N} \dfrac{\Lambda_{ij} (r \mathcal{N}_j)}{r} \right) \,.
\end{equation}
It is also assumed that the diagonal terms satisfy $\Lambda_{ii}=1$, which corresponds to the \emph{intraspecific competition} of species $i$ in the ecological network. After this rescaling, and setting $r=1$, the dynamics can be written in terms of the dimensionless gLV equation
\begin{equation}
	\label{eq:GLV}
	\dfrac{d \mathcal{N}_i}{dt} = \mathcal{N}_i \left(1 - \mathcal{N}_i - \sum_{\substack{j=1\\ j\neq i}}^{N} \Lambda_{ij}\,\mathcal{N}_j \right).
\end{equation}
In this formulation, $\mathcal{N}_i(t)$ represents the rescaled abundance of species $i$, related to the original variables by $x_i = r \mathcal{N}_i$. In the remainder of the manuscript we use $\mathcal{N}_i(t)$ consistently to denote species abundances in the dimensionless gLV dynamics defined by Eq. (\ref{eq:GLV}).
\subsection{Dissimilarity between two gLV processes on networks} 
In this section, we study the evolution of $\mathcal{N}_1(t), \ldots, \mathcal{N}_N(t)$ governed by Eq.~(\ref{eq:GLV}) on a network with interactions given by $\bm{\Lambda}$, referred to as the reference dynamics. We also analyze a modified process, $\mathcal{N}^\star_1(t), \ldots, \mathcal{N}^\star_N(t)$, which follows the same equation but with a different interaction matrix $\bm{\Lambda}^\star$. Both processes start from identical initial conditions, $\mathcal{N}_i(0) = \mathcal{N}^\star_i(0)$ for $i=1,\dots,N$. To compare them, we define the dissimilarity
\begin{equation}
	\label{eq:Def_Dist_Red_genLV}
	\mathcal{D}^{(\mathrm{LV})}(t) \equiv \frac{1}{\mathcal{N}_0} \sum_{i=1}^{N} \left| \mathcal{N}_i(t) - \mathcal{N}^\star_i(t) \right|,
\end{equation}
where $\mathcal{N}_0 = \sum_{i=1}^{N} \mathcal{N}_i(0) = \sum_{i=1}^{N} \mathcal{N}^\star_i(0)$. This expression generalizes the two-species definition in Eq.~(\ref{eq:D(t)}) to systems with $N$ interacting species. The normalization by the total initial population $\mathcal{N}_0$ makes the quantity dimensionless and allows it to be interpreted as the fraction of the initial population separating the two dynamical states at time $t$. In this way, the dissimilarity has a direct interpretation in terms of relative population deviations, which facilitates comparisons across systems with different initial population sizes.
\\[2mm]
Alternative distance measures could also be considered for comparing population trajectories. For instance, Euclidean distances in the space of population vectors provide a mathematically valid metric. However, such measures tend to emphasize large deviations in a small number of species and obscure the direct interpretation in terms of aggregate population differences. In contrast, the definition in Eq.~(\ref{eq:Def_Dist_Red_genLV}) treats deviations across species in a linear and additive manner, which provides a more transparent interpretation in terms of total population differences between the two processes. Similarly, information--theoretic divergences are typically defined for probability distributions, whereas the variables $\mathcal{N}_i(t)$ represent population abundances rather than normalized densities, making such formulations less natural in the present context.
\\[2mm]
To summarize the overall difference between two processes, it is also useful to define the maximum dissimilarity
\begin{equation}
	\mathcal{D}_{\mathrm{max}}\{\bm{\Lambda}, \bm{\Lambda}^\star\} \equiv \max_{t' \geq 0} \mathcal{D}^{(\mathrm{LV})}(t'),
	\label{eq:dmax_red}
\end{equation}
which condenses the entire temporal evolution of $\mathcal{D}^{(\mathrm{LV})}(t)$ into a single scalar quantity characterizing the largest separation between the two dynamical trajectories.
\\[2mm]
Our motivation for introducing these dissimilarity measures is to provide a systematic framework for comparing dynamical processes that may differ either in their interaction structure or in the equations governing their evolution. In this context, a suitable measure should satisfy several desirable properties. First, it should be sensitive to modifications in the interaction matrix, regardless of whether such changes are localized in a small subset of links or distributed across the entire network. Second, it should detect differences arising from changes in the functional form of the governing equations, thereby enabling comparisons between processes that do not necessarily follow identical dynamical rules. Third, the measure should reflect the magnitude of the dynamical divergence in a proportional manner, such that larger differences in the evolution of the populations correspond to larger values of the dissimilarity. Finally, it should retain temporal information about the dynamics, allowing the identification of when the divergence between the processes emerges and whether these differences persist or vanish in the long-time stationary regime.
\\[2mm]
It is also important to note that when the interaction structure of the modified system leads to unstable dynamics, population abundances may grow without bound. In such cases the quantity $\mathcal{D}^{(\mathrm{LV})}(t)$ may also diverge. Rather than limiting the usefulness of the measure, this behavior can also serve as an indicator that the modified system has entered an unstable dynamical regime. As shown in the Results section, the divergence of $\mathcal{D}^{(\mathrm{LV})}(t)$ can therefore signal transitions from stable dynamics to unstable population growth induced by changes in the interaction structure. Taken together, the quantities introduced in Eqs.~(\ref{eq:Def_Dist_Red_genLV}) and (\ref{eq:dmax_red}) provide a flexible framework for comparing nonlinear dynamical processes on networks.
\\[2mm]
Finally, we emphasize that the simplifying assumptions adopted, such as considering identical intrinsic growth rates across species in Eq. (\ref{eq:GLV}), were introduced to isolate the effects of interaction structure and to enable a clearer interpretation of the resulting dynamics. These assumptions are not inherent limitations of the proposed framework. The dissimilarity measures defined in Eqs.~(\ref{eq:Def_Dist_Red_genLV}) and (\ref{eq:dmax_red}) depend only on the comparison of dynamical processes and can be readily extended to more general settings, including systems with heterogeneous growth rates, alternative functional responses, and additional ecological mechanisms. Accordingly, the framework is not restricted to the specific formulations considered here, but provides a flexible and broadly applicable approach for investigating how structural and parametric variations shape the dynamics of complex interacting systems.
\section{Results}
In this section, we present numerical results illustrating how the proposed dissimilarity measures capture differences between dynamical processes described by generalized Lotka--Volterra equations. The analysis is organized in a progressive manner, beginning with the minimal case of two interacting species, where the effect of parameter variations can be directly interpreted in terms of predator--prey oscillations. We then extend the framework to systems defined on small directed networks, allowing us to analyze how changes in interaction strengths or network topology influence the resulting dynamics. Building on these examples, we investigate larger systems with community structure, where interactions are distributed across cliques and random perturbations can affect stability and global behavior. Finally, we demonstrate that the proposed methodology also enables the comparison of processes governed by different nonlinear dynamical equations, highlighting the generality of the framework. This progression from simple to increasingly complex scenarios illustrates how dynamical dissimilarity measures provide a systematic way to detect and quantify differences across a wide range of interacting systems.

\subsection{Dissimilarity in systems with two species }
\begin{figure}[b!]
	\centering
	\includegraphics*[width=1.0\textwidth]{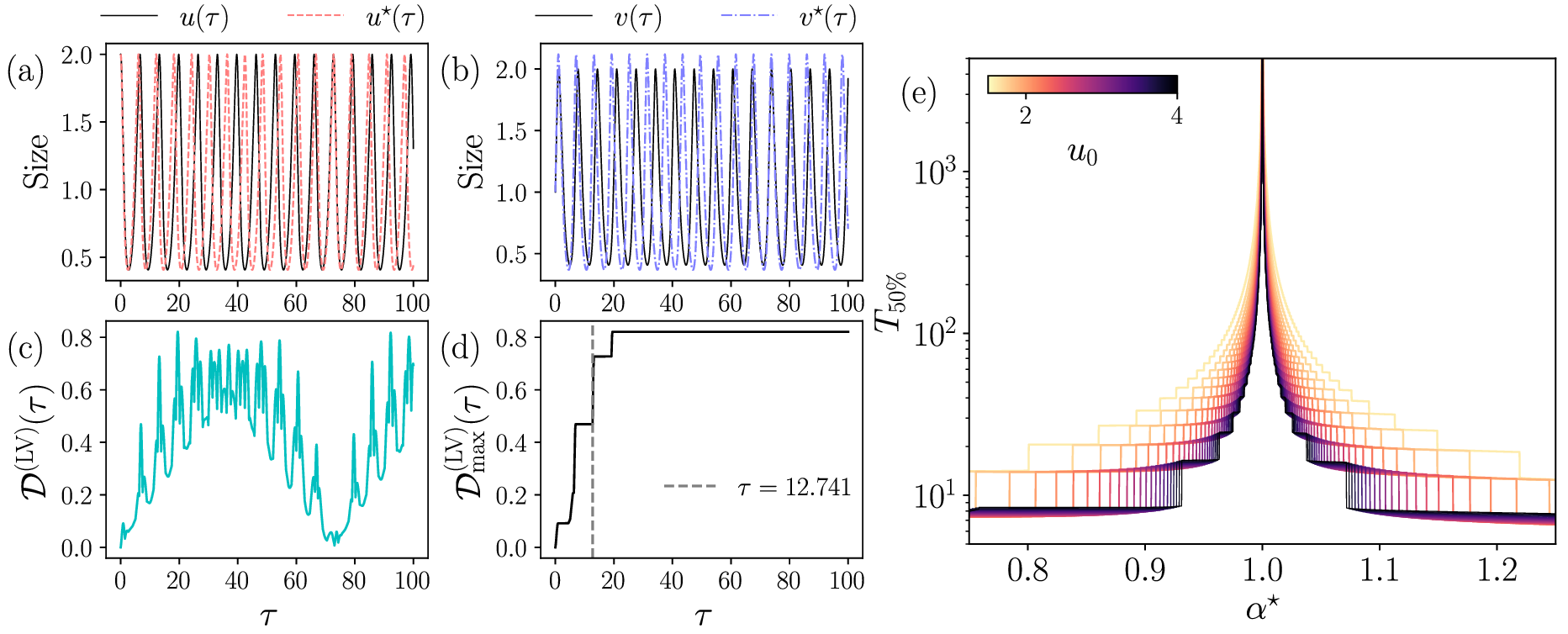}
	\vspace{-2mm}
	\caption{Dissimilarity  in the dynamics of  prey and predator populations using a reference system with $\alpha = 1$. (a) Functions $u(\tau)$ (original process) and $u^\star(\tau)$ (modified process with $\alpha^\star = 1.2$) associated to preys as a function of $\tau$, with $u_{0} = u_{0}^{\star} = 2$, panel (b) shows the same result for predators $v(\tau)$ and $v^\star(\tau)$. (c) Dissimilarity metric $\mathcal{D}^{(\mathrm{LV})}(\tau)$ in Eq. (\ref{eq:D(t)}) for the two systems in (a)–(b), $\{(u,v), (u^{\star}, v^{\star})\}$, as a function of time. (d) Maximum values $\mathcal{D}^{(\mathrm{LV})}_{\mathrm{max}}(\tau)$; the time $\tau = 12.741$ corresponds to the first moment when the two systems exhibit a $50\%$ difference with respect to the initial conditions. (e) Time $T_{50\%}$ required for two systems (reference–modified) to differ by $50\%$ as a function of $\alpha^\star$, for different values of $u_0 = u(0)$ (indicated by the colors of each line).  }
	\label{Fig_1}
\end{figure}
Figure \ref{Fig_1} presents a numerical example for the analysis of dissimilarity in systems with two interacting species, as defined in Section \ref{Sec_LVmodel}, evolving according to Eqs. (\ref{eq:LVA2}). The theoretical framework developed in Section \ref{Sec_diss_LV2} is applied to a reference system with $\alpha = 1$ and to a modified system with $\alpha^{\star} = 1.2$. Figure  \ref{Fig_1}(a) shows the temporal evolution of the prey populations for both the reference $u(\tau)$ and the modified $u^\star(\tau)$ systems, starting from the same initial condition $u_{0} = u^{\star}_{0} = 2$. The modified system reaches its first maximum in a shorter time than the reference system, indicating a higher oscillation frequency of the populations. A similar pattern is observed in Fig. \ref{Fig_1}(b), which displays the predator populations for the $v(\tau)$ and the modified $v^\star(\tau)$ systems, both starting from the initial condition $v_{0} = v^{\star}_{0} = 1$.
\\[2mm]
Since the two systems are characterized by different values of the interaction parameter, their oscillation frequencies are also different. As shown in Section~\ref{Sec_diss_LV2}, the linear stability analysis of the coexistence equilibrium yields a characteristic frequency $\omega=\sqrt{\alpha}$ for the population oscillations. Consequently, the reference system with $\alpha=1$ and the modified system with $\alpha^\star=1.2$ evolve with different frequencies, leading to a gradual phase mismatch between their trajectories. As a result, there are times at which the populations of the two systems nearly coincide, whereas at other times one system is close to a maximum while the other is close to a minimum. These differences between the respective populations are particularly evident at early times. Figure \ref{Fig_1}(c) shows $\mathcal{D}^{(\mathrm{LV})}(\tau)$, defined in Eq.~(\ref{eq:D(t)}), as a function of $\tau$. The results exhibit an oscillatory behavior characteristic of this quantity for the case of two interacting species. This behavior arises from the phase difference between the two systems induced by their different oscillation frequencies. In particular, the modulation pattern observed in $\mathcal{D}^{(\mathrm{LV})}(\tau)$ reflects the progressive accumulation of phase differences between oscillations with frequencies $\sqrt{\alpha}$ and $\sqrt{\alpha^\star}$, which generates a beating effect governed by the frequency difference $\sqrt{\alpha^\star}-\sqrt{\alpha}$. It is also important to analyze the maxima of the calculated differences, $\mathcal{D}^{(\mathrm{LV})}_{\mathrm{max}}(\tau)$, given by Eq.~(\ref{eq:dmax}), as a function of the elapsed time $\tau$. Figure \ref{Fig_1}(d) depicts $\mathcal{D}^{(\mathrm{LV})}_{\mathrm{max}}(\tau)$. For the systems studied under the previously described conditions, the results show that the time at which the reference system differs from the modified one by $50\%$ is relatively short, $\tau = 12.741$, compared with the maximum time $\tau_{\max}=100$.
\\[2mm]
The results shown in Figs. \ref{Fig_1}(a)–(d), which characterize a Lotka--Volterra-type predator–prey system $(u_{0}, v_{0}, \alpha)$, highlight that the parameter $\alpha$ modulates the coupling between species, thus explicitly determining the system's evolution. The results in Figs. \ref{Fig_1}(c)–(d) suggest that an alternative way to quantify the differences between two-species systems is to consider the minimum time at which a difference of $\nu\%$ with respect to the initial populations is observed. This time, denoted $T_{\nu\%}$, is obtained as the minimum $\tau$ for which $\mathcal{D}^{(\mathrm{LV})}_{\mathrm{max}}(\tau) \geq \nu/100$.
\\[2mm]
Figure \ref{Fig_1}(e) illustrates the dependence of $T_{50\%}$, corresponding to a dissimilarity level of $\nu\% = 50\%$, on the parameter $\alpha^\star$, for different initial prey populations $u_0$ (encoded in the color bar). The results show that as $\alpha^\star \approx 1$, the time required to reach a $50\%$ dissimilarity tends to increase, eventually reaching the maximum simulation time. In the figure, each curve corresponds to a different $u_{0}$, with $1.6 \le u_{0} \le 4$, and is represented by a distinct color and line width. The predator population was fixed at $v_{0} = 1$ for all realizations.
\\[2mm]
The results in Fig. \ref{Fig_1}(e) provide numerical evidence supporting the hypothesis that increasing the initial number of prey–predator individuals leads to earlier manifestation of differences between the systems. This indicates that the dynamics are markedly distinct when starting from larger populations, in contrast to the behavior observed for smaller populations. This observation is relevant in the context of complex systems, particularly in ecological and biological settings \cite{may1972,Liu2023,Akjouj2024}. The analysis of the dimensionless two-species Lotka--Volterra-type model reveals that prey and predator populations exhibit oscillatory behaviors that are highly sensitive to initial conditions and to the interaction parameter $\alpha$. Numerical simulations comparing a reference system with $\alpha = 1$ and a modified system with $\alpha^{\star} = 1.2$ show differences in the amplitude and frequency of oscillations around the stable equilibrium point. The dissimilarity metric $\mathcal{D}^{(\mathrm{LV})}(\tau)$ quantifies these differences over time, showing that the populations of the two systems diverge significantly, especially when the parameter $\alpha^\star$ in the modified process is increased. It was found that the time $T_{50\%}$ required for the differences between the two systems to reach a $50\%$ dissimilarity decreases with larger values of $\alpha^\star$ and with higher initial prey populations.
\\[2mm]
Our findings reported in Fig. \ref{Fig_1} emphasize the importance of the interaction parameter in the dynamics of ecological systems, suggesting that variations in interspecies interactions can lead to markedly different behaviors. This has direct implications for understanding population dynamics in more complex biological systems.
\subsection{Generalized Lotka–Volterra model in networks with three nodes}
\begin{figure}[t!]
	\centering
	\includegraphics*[width=1.0\textwidth]{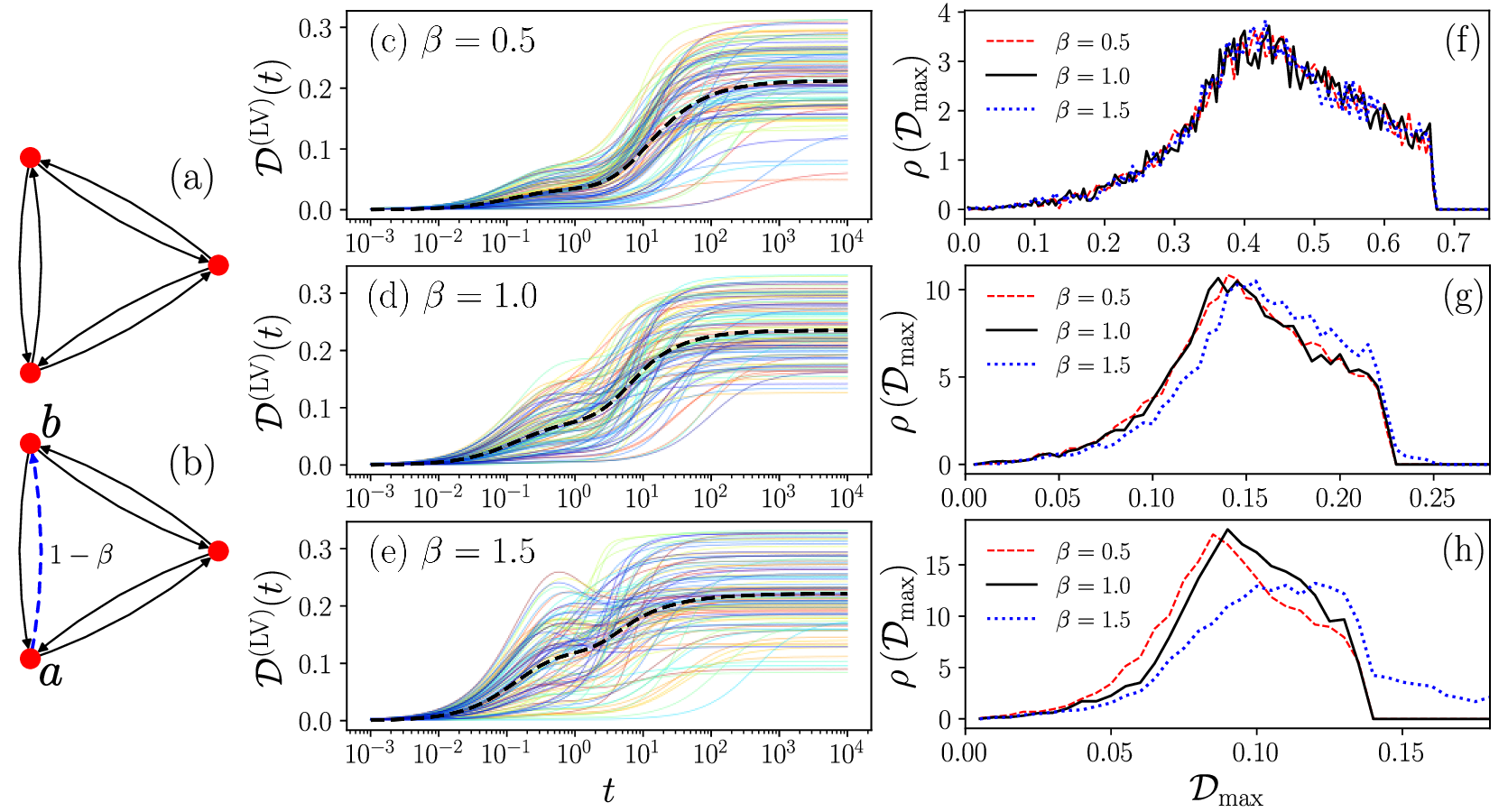}
	\vspace{-2mm}
	\caption{Comparison of two gLV dynamical processes in a graph with three nodes.  
		(a) Reference case: a complete graph with uniform interactions, $\Lambda_{ij} = 1$.  
		(b) Modified graph: a single interaction is altered with $\Lambda^\star_{ab} = 1 - \beta$ for the link $a \to b$, where $\beta \geq 0$; all other entries $\Lambda^\star_{ij}$ remain as in the reference. (c)-(e) Temporal evolution of the dissimilarity $\mathcal{D}^{(\mathrm{LV})}(t)$ between the dynamics on the reference graph (a) and the modified graph for $\beta=0.5$, $1.0$, and $1.5$, respectively. The curves correspond to $100$ realizations with random initial conditions satisfying $\mathcal{N}_0=2$, and the dashed line represents the average over all realizations $\langle\mathcal{D}^{(\mathrm{LV})}(t)\rangle$.  (f) Probability density $\rho(\mathcal{D}_{\mathrm{max}})$ of the maximum dissimilarity $\mathcal{D}_{\mathrm{max}}=\mathcal{D}_{\mathrm{max}}(\bm{\Lambda},\bm{\Lambda}^\star)$ between the two structures for different values of $\beta$ and initial population values $\mathcal{N}_0$. The results are obtained from histogram counts using a bin size $\Delta \mathcal{D}_{\mathrm{max}} = 5\times10^{-3}$.  The results in (f) correspond to $\mathcal{N}_0 = 1$, using random initial conditions and $10^4$ realizations of the process.  
		The same analysis is repeated in (g) for $\mathcal{N}_0 = 3$, and in (h) for $ \mathcal{N}_0 = 5$.
	}
	\label{Fig_2}
\end{figure}

\begin{figure}[t!]
	\centering
	\includegraphics*[width=0.95\textwidth]{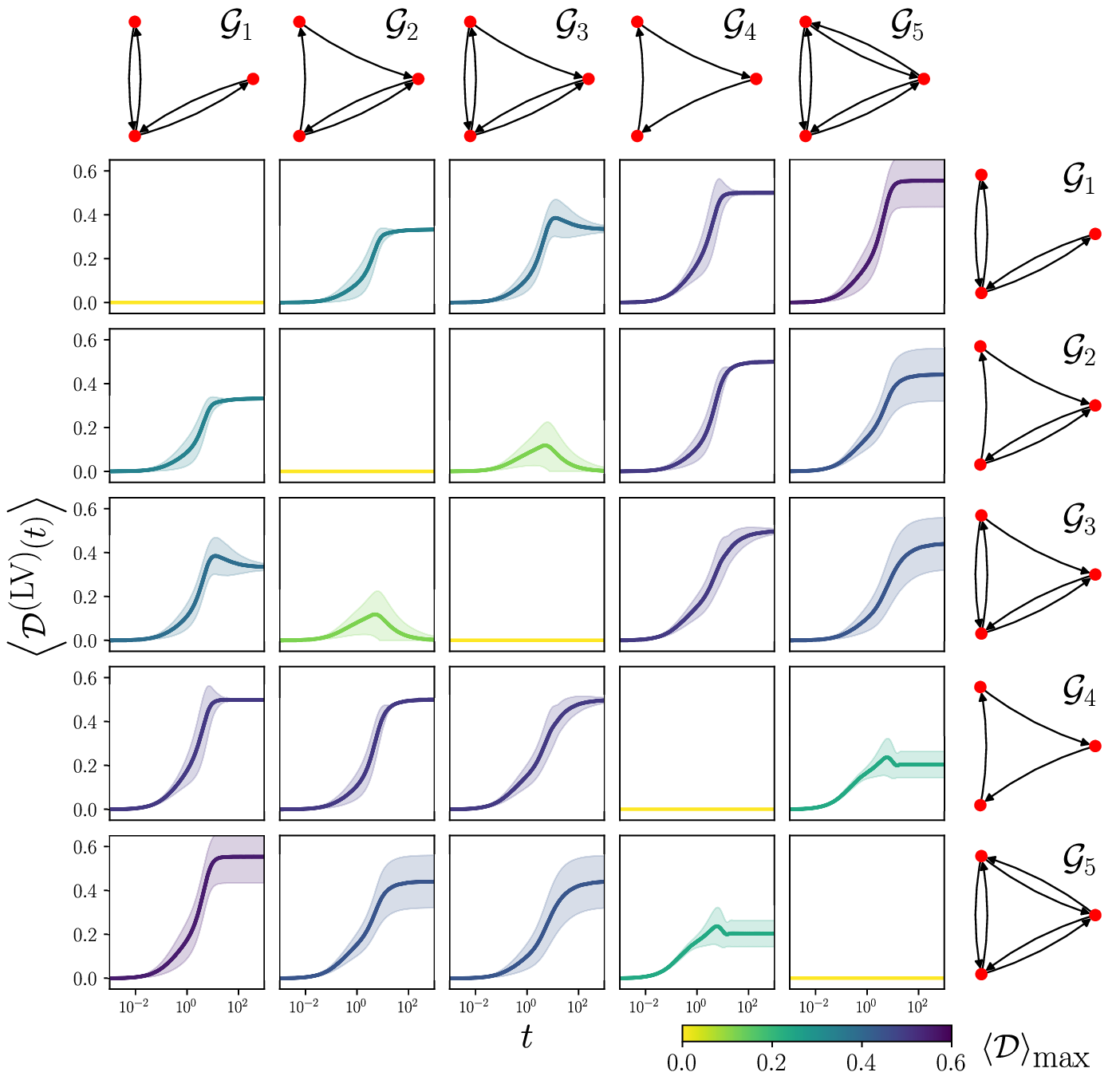}
	\vspace{1mm}
	\caption{Comparison of gLV dynamical processes for all non-isomorphic connected directed graphs of size $N = 3$.  The thick continuous lines show the ensemble average $\left\langle \mathcal{D}^{(\mathrm{LV})}(t) \right\rangle$ as a function of time $t$, computed from $10^4$ realizations with random initial conditions fulfilling  $\mathcal{N}_0 = 1$.  Results are arranged in panels where row $\mu$ and column $\nu$ correspond to the comparison between the dynamics defined by the graph $\mathcal{G}_\mu$ and the graph $\mathcal{G}_\nu$.  For each time $t$, the shaded region around the average represents the standard deviation of $\mathcal{D}^{(\mathrm{LV})}(t)$ obtained across different realizations. In the comparison between $\mathcal{G}_2$ and $\mathcal{G}_3$, the shaded region is restricted to positive values only. The colors, encoded in the colorbar, represent the maximum value $\left\langle \mathcal{D}\right\rangle_\mathrm{max}$ of the average results.}
	\label{Fig_3}
\end{figure}

In this section, we compare two complete networks that share the same topology but differ in a single element of the interaction matrix. The reference network is defined by $\Lambda_{ij}=1$ for $i \neq j$, as shown in Fig. \ref{Fig_2}(a). In the modified network, only the interaction along the link $a \to b$ is changed such that $\Lambda_{ab}^{\star}=1-\beta$, with $0 \leq \beta \leq 2$, as illustrated by the dashed link in Fig. \ref{Fig_2}(b). This range ensures that the modified interaction $\Lambda_{ab}^{\star}=1-\beta$ spans positive, null, and negative values. In particular, $\beta=0$ corresponds to the reference case, $\beta=1$ removes the interaction, and $\beta>1$ induces a sign change. Thus, the reference network corresponds to a fully connected network with competitive interactions, whereas the modified network contains a single link whose weight varies as a function of $\beta$. Figures \ref{Fig_2}(c)–(e) show the dissimilarity $\mathcal{D}^{(\mathrm{LV})}(t)$ as a function of time between the dynamics on the two networks. The systems are initialized with random initial conditions satisfying $\mathcal{N}_0=2$, and we consider the representative values $\beta=0.5$, $1.0$, and $1.5$, which illustrate three qualitatively distinct regimes. In all cases, the dashed line represents the average over realizations $\langle\mathcal{D}^{(\mathrm{LV})}(t)\rangle$. The results show that for all realizations $\mathcal{D}^{(\mathrm{LV})}(t \to 0)=0$, after which the dissimilarity increases and eventually stabilizes at a constant value at long times. However, the growth of the curves occurs, on average, over different time scales, and the characteristic time at which observable changes in magnitude emerge decreases as $\beta$ increases. This result indicates that larger values of $\beta$ enable a faster detection that the reference system has been modified.
\\[2mm]
In the cases analyzed in Figs. \ref{Fig_2}(c)-(e), it is worth noting that the maximum values of $\mathcal{D}_{\mathrm{max}}(\bm{\Lambda},\bm{\Lambda}^\star)$ are similar for the values of $\beta$ explored. However, this behavior may change as the average of the initial conditions, $\mathcal{N}_0$, increases due to the nonlinear nature of the system. This observation motivates a statistical examination of $\mathcal{D}_{\mathrm{max}}(\bm{\Lambda},\bm{\Lambda}^\star)$ over a larger number of realizations in order to assess how $\mathcal{N}_0$ influences this quantity. To characterize the statistical variability of the quantities obtained from the numerical realizations, we compute the probability density of the corresponding values. In this context, a probability density $\rho(x)$ describes how frequently a variable $x$ takes values within a given interval when many realizations of the system are considered. More precisely, the probability that the variable lies between $x$ and $x+\Delta x$ is given by
$P(x \leq X < x+\Delta x) = \rho(x)\,\Delta x$. The function $\rho(x)$ is normalized such that
$\int \rho(x)\,dx = 1$, which ensures that all possible outcomes are accounted for. In practice, $\rho(x)$ is estimated from the data by constructing a histogram of the values obtained from many realizations and normalizing it so that the total area equals one.
\\[2mm]
In Figs. \ref{Fig_2}(f)-(h), we perform a statistical analysis of the probability density $\rho(\mathcal{D}_{\mathrm{max}})$ of the maximum dissimilarity, denoted as $\mathcal{D}_{\mathrm{max}} = \mathcal{D}_{\mathrm{max}}\{\bm{\Lambda}, \bm{\Lambda}^\star\}$. Figure \ref{Fig_2}(f) shows $\rho(\mathcal{D}_{\mathrm{max}})$ for $\beta = 0.5,\, 1.0,$ and $1.5$, computed over $10^4$ realizations with random initial conditions satisfying $\mathcal{N}_0 = 1$. The same statistical analysis is repeated in Figs. \ref{Fig_2}(g)-(h) for $\mathcal{N}_0 = 3$ and $\mathcal{N}_0 = 5$. The results indicate that $\rho(\mathcal{D}_{\mathrm{max}})$ varies very little across the analyzed values of $\beta$ when $\mathcal{N}_0 = 1$ [see Fig. \ref{Fig_2}(f)]. In contrast, when the initial conditions are changed to $\mathcal{N}_0 = 3$ or $\mathcal{N}_0 = 5$, $\rho(\mathcal{D}_{\mathrm{max}})$ exhibits clear dependence on $\beta$ [see Figs. \ref{Fig_2}(g)-(h)]. This result highlights the role of nonlinearity in the system, showing that when comparing the two networks it is necessary to also account for the initial population conditions.
\\[3mm]
In Fig. \ref{Fig_2}, we examined dissimilarities using a reference network and assessed the effect of weight modifications in this structure. The proposed approach is, however, general and can be applied to detect differences in gLV dynamics between any pair of connected networks of the same size. To illustrate this, we compare the dynamics of all non-isomorphic, connected, directed networks with $N=3$ nodes. These graphs, denoted by $\mathcal{G}_1,\, \mathcal{G}_2,\ldots, \mathcal{G}_5$, are shown in Fig. \ref{Fig_3}. They represent several topologies of interest: $\mathcal{G}_1$ is a linear graph, $\mathcal{G}_4$ is a directed cycle, and $\mathcal{G}_5$ is a triangle.  
\\[2mm]
In Fig. \ref{Fig_3}, we compare the dynamics considering pairs of graphs $\mathcal{G}_\mu$ and $\mathcal{G}_\nu$ with $\mu,\nu=1,2,\ldots,5$, where $\mu$ indexes the rows and $\nu$ the columns of the panels. The graphs under comparison are displayed along the top and right borders. For each case, the interaction matrix $\bm{\Lambda}$ is derived from the adjacency matrix $\bm{A}$ of the corresponding graph, with elements $A_{ij}=1$ if a directed link exists from node $i$ to node $j$, and $A_{ij}=0$ otherwise.   Each panel in row $\mu$ and column $\nu$ compares the dynamics defined by the reference graph $\mathcal{G}_\mu$ with those of the modified dynamics defined by $\mathcal{G}_\nu$. The results are obtained by numerically integrating Eq. (\ref{eq:GLV}) with random initial conditions $\mathcal{N}_i(0)=\mathcal{N}_i^\star(0)$, subject to the constraint $\mathcal{N}_0=\sum_{i=1}^3 \mathcal{N}_i(0)=1$. The curves represent the ensemble average $\langle \mathcal{D}^{(\mathrm{LV})}(t) \rangle$ as a function of time, computed over $10^4$ realizations of the initial conditions. Shaded regions indicate the standard deviation at each $t$. The comparison of all graph pairs produces 25 curves, each colored according to the corresponding maximum value of $\langle \mathcal{D}^{(\mathrm{LV})}(t) \rangle$, denoted as $\langle \mathcal{D}\rangle_{\mathrm{max}}$, and encoded in the color bar.  
\\[2mm]
The results show that $\langle \mathcal{D}^{(\mathrm{LV})}(t) \rangle = 0$ for the panels on the diagonal, since in these cases the dynamics are compared using identical interaction matrices. In the other cases, the curves $\langle \mathcal{D}^{(\mathrm{LV})}(t) \rangle$ start from zero due to the identical initial conditions, but then increase significantly before reaching a stationary value. This behavior is generally observed, except for the comparison between graphs $\mathcal{G}_2$ and $\mathcal{G}_3$, where $\langle \mathcal{D}^{(\mathrm{LV})}(t \to \infty) \rangle = 0$. Notably, for the comparisons between $\mathcal{G}_1$ and $\mathcal{G}_3$, and between $\mathcal{G}_4$ and $\mathcal{G}_5$, the $\langle \mathcal{D}^{(\mathrm{LV})}(t) \rangle$ first increases to a maximum value and then decreases to a nonzero stationary value.
\\[2mm]
On the other hand, the numerical values of $\langle \mathcal{D}\rangle_{\mathrm{max}}$ provide a single quantitative measure of the differences between two dynamical processes. For example, the panels in the first row of Fig. \ref{Fig_3} compare the gLV dynamics using $\mathcal{G}_1$ as the reference. The values of $\langle \mathcal{D}\rangle_{\mathrm{max}}$ reveal that the most similar process occurs in $\mathcal{G}_2$, and that the dissimilarity gradually increases for the other graphs, reaching its maximum with the complete graph $\mathcal{G}_5$.  In the second row, when $\mathcal{G}_2$ is used as a reference, the dynamics is most similar to that on $\mathcal{G}_3$, which differs from $\mathcal{G}_2$ only by the presence of one additional edge. In contrast, when $\mathcal{G}_3$ is taken as the reference (third row), the closest dynamics are found in $\mathcal{G}_2$ and $\mathcal{G}_1$, which differ from $\mathcal{G}_3$ by the removal or addition of a single directed edge.  The analyses in the fourth row, using the ring $\mathcal{G}_4$ as a reference, show that the most similar process occurs in $\mathcal{G}_5$, while the largest dissimilarity $\langle \mathcal{D}\rangle_{\mathrm{max}}$ is obtained when $\mathcal{G}_4$ is compared with $\mathcal{G}_{1}$. Finally, when the fully connected graph $\mathcal{G}_5$ is used as reference (fifth row in Fig. \ref{Fig_3}), the values of $\langle \mathcal{D}\rangle_{\mathrm{max}}$ decrease monotonically from $\mathcal{G}_1$ to $\mathcal{G}_5$, indicating that the largest difference arises between the dynamics on $\mathcal{G}_5$ and the linear graph $\mathcal{G}_1$. These results demonstrate that the proposed approach effectively detects and quantifies differences in the gLV dynamics on networks. For example, if one considers only the entries of interaction matrices (which for non-diagonal elements in this case coincide with the adjacency matrices), it is impossible to determine whether the process on $\mathcal{G}_1$ or that on $\mathcal{G}_2$ is more similar to the dynamics on $\mathcal{G}_5$, since both $\mathcal{G}_1$ and $\mathcal{G}_2$ contain the same number of links and satisfy $\sum_{i,j}(\Lambda^\star_{ij}-\Lambda_{ij})=2$. Nevertheless, the results clearly show that, between $\mathcal{G}_1$ and $\mathcal{G}_2$, the dynamics on $\mathcal{G}_2$ is more similar to that on $\mathcal{G}_5$.
\\[2mm]
It is important to emphasize that the magnitude of $\langle \mathcal{D} \rangle_{\mathrm{max}}$ is not determined solely by the number of edges that differ between two graphs. For example, the comparison between $\mathcal{G}_3$ and $\mathcal{G}_5$ yields a larger value of $\langle \mathcal{D} \rangle_{\mathrm{max}}$ than the comparison between $\mathcal{G}_4$ and $\mathcal{G}_5$, even though the latter pair differs by a larger number of edges. This occurs because the gLV dynamics depend not only on the presence or absence of interactions, but also on how these interactions participate in feedback loops that determine the stability and transient evolution of the system. In nonlinear dynamical systems, a single modification in the interaction structure can alter these feedback mechanisms and produce a larger deviation between trajectories than multiple structural changes that preserve the qualitative stability properties of the network. Therefore, the values of $\langle \mathcal{D} \rangle_{\mathrm{max}}$ capture the dynamical consequences of structural variations rather than simply counting topological differences between graphs.
\\[2mm]
\begin{figure}[t]
	\centering
	\includegraphics*[width=0.95\textwidth]{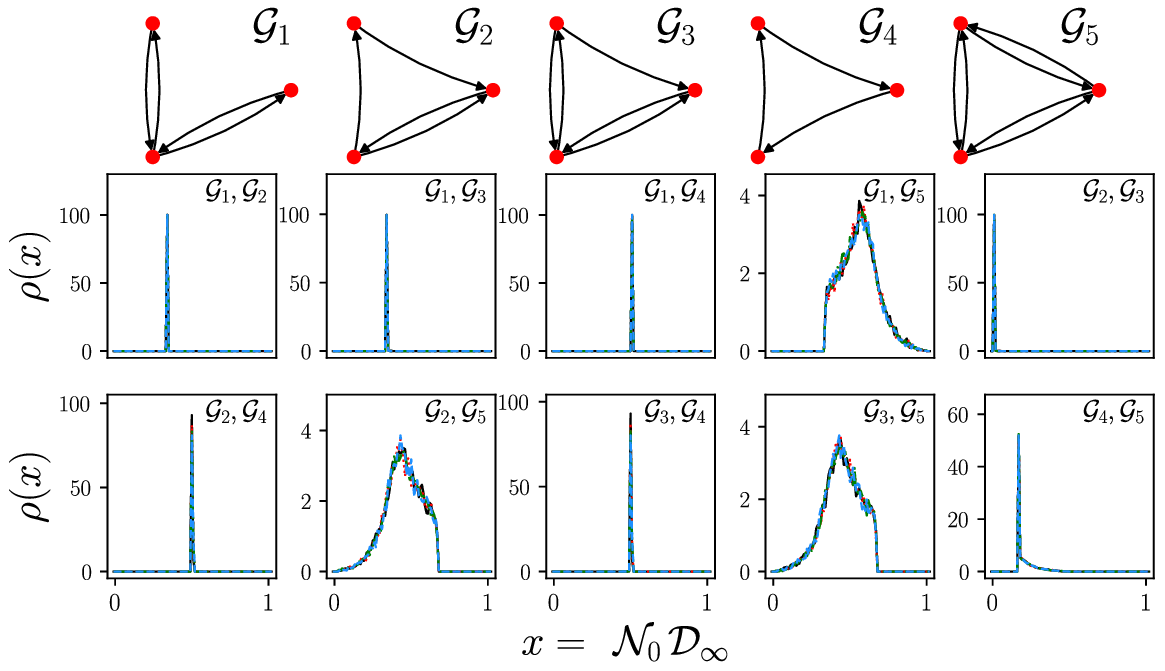}
	\vspace{0mm}
	\caption{Statistical analysis of $\mathcal{D}_\infty$ for the comparison of gLV dynamical processes on graphs of size $N=3$. The panels show the probability density $\rho(x)$ of the rescaled variable $x=\mathcal{N}_0 \mathcal{D}_\infty$, obtained from comparisons between the dynamics on pairs of graphs $\mathcal{G}_1, \mathcal{G}_2, \ldots, \mathcal{G}_5$. The results are computed from $10^4$ realizations with random initial conditions for $\mathcal{N}_0=1$, $2$, $3$, and $4$. Histogram counts are obtained using a bin size $\Delta x = 10^{-2}$.}
	\label{Fig_4}
\end{figure}
The results in Figs. \ref{Fig_2} and \ref{Fig_3} show that, in addition to the curves $\langle \mathcal{D}^{(\mathrm{LV})}(t) \rangle$ and their maxima, it is also important to consider the value $\mathcal{D}_\infty \equiv \lim_{t \to \infty} \mathcal{D}^{(\mathrm{LV})}(t)$ for each realization of the initial conditions. In the following analysis, we do not restrict the discussion to the ensemble average of this quantity, but instead examine the statistical properties of the individual values obtained. With this purpose, Fig. \ref{Fig_4} presents the probability density $\rho(x)$ of the values $x = \mathcal{N}_0 \mathcal{D}_\infty$ obtained from the comparison of all pairs of different non-isomorphic, connected, directed networks with $N=3$ nodes. These networks define the evolution of the gLV dynamics in Eq. (\ref{eq:GLV}), where the interaction matrices are determined by the adjacency matrix of each graph. The ten panels in Fig. \ref{Fig_4} show the results for the pairs of graphs $\mathcal{G}_\mu$, $\mathcal{G}_\nu$ under analysis.
\\[2mm]
As a first result, Fig. \ref{Fig_4} shows that for the comparisons $\mathcal{G}_1$ with $\mathcal{G}_2$, $\mathcal{G}_1$ with $\mathcal{G}_3$, $\mathcal{G}_1$ with $\mathcal{G}_4$, $\mathcal{G}_2$ with $\mathcal{G}_4$, and $\mathcal{G}_3$ with $\mathcal{G}_4$, the probability densities reveal that the values of $\mathcal{N}_0 \mathcal{D}_\infty$ are concentrated in a sharp peak, resembling a Dirac delta. A particularly notable case is the pair $\mathcal{G}_2$ and $\mathcal{G}_3$, for which the average $\langle \mathcal{D}_\infty \rangle = 0$. This result indicates that, at long times, the corresponding systems reach the same stationary equilibrium, and differences between the two dynamics are observed only during the transient regime, as reflected in the curve $\langle \mathcal{D}^{(\mathrm{LV})}(t) \rangle$. Such behavior cannot be inferred solely from the topological structure of the two graphs.
\\[2mm]
On the other hand, when comparing $\mathcal{G}_1$ with $\mathcal{G}_5$, the probability density $\rho(x)$ vanishes for $x \leq 0.3$. For larger values of $x$, $\rho(x)$ increases rapidly, reaches a maximum, and then decays gradually until $x=1$. In contrast, for the comparisons $\mathcal{G}_2$ with $\mathcal{G}_5$ and $\mathcal{G}_3$ with $\mathcal{G}_5$, the opposite behavior is observed: $\rho(x)$ increases gradually from $x=0$, reaches a maximum, and then decays to $0$ at $x \approx 0.7$, followed by an interval where $\rho(x)=0$ up to $x=1$. In addition, for the pair $\mathcal{G}_4$ and $\mathcal{G}_5$, the results show that many values of $x$ are concentrated around the same result, with only a small fraction deviating slightly from the peak value. Finally, it is worth noting that, for the cases analyzed, the curves of $\rho(x)$ are similar for the values $\mathcal{N}_0 = 1, 2, 3,$ and $4$ considered.
\\[2mm]
The analysis of three-node graphs presented in this section illustrates different ways to compare gLV processes on networks under random initial conditions. The first and most relevant measure is $\left\langle \mathcal{D}^{(\mathrm{LV})}(t) \right\rangle$, which reveals when the two systems start to diverge, highlights intervals with transient deviations, and shows how these stabilize into stationary values. The maximum of this curve, $\left\langle \mathcal{D}\right\rangle_\mathrm{max}$, identifies the largest separation and provides a compact quantification with a single number. Finally, a statistical analysis of $\mathcal{D}_\infty$ offers insight into whether the equilibrium states converge to the same or to distinct values.
\subsection{Dissimilarity of random networks with cliques} 
\begin{figure}[t!]
	\centering
	\includegraphics*[width=0.95\textwidth]{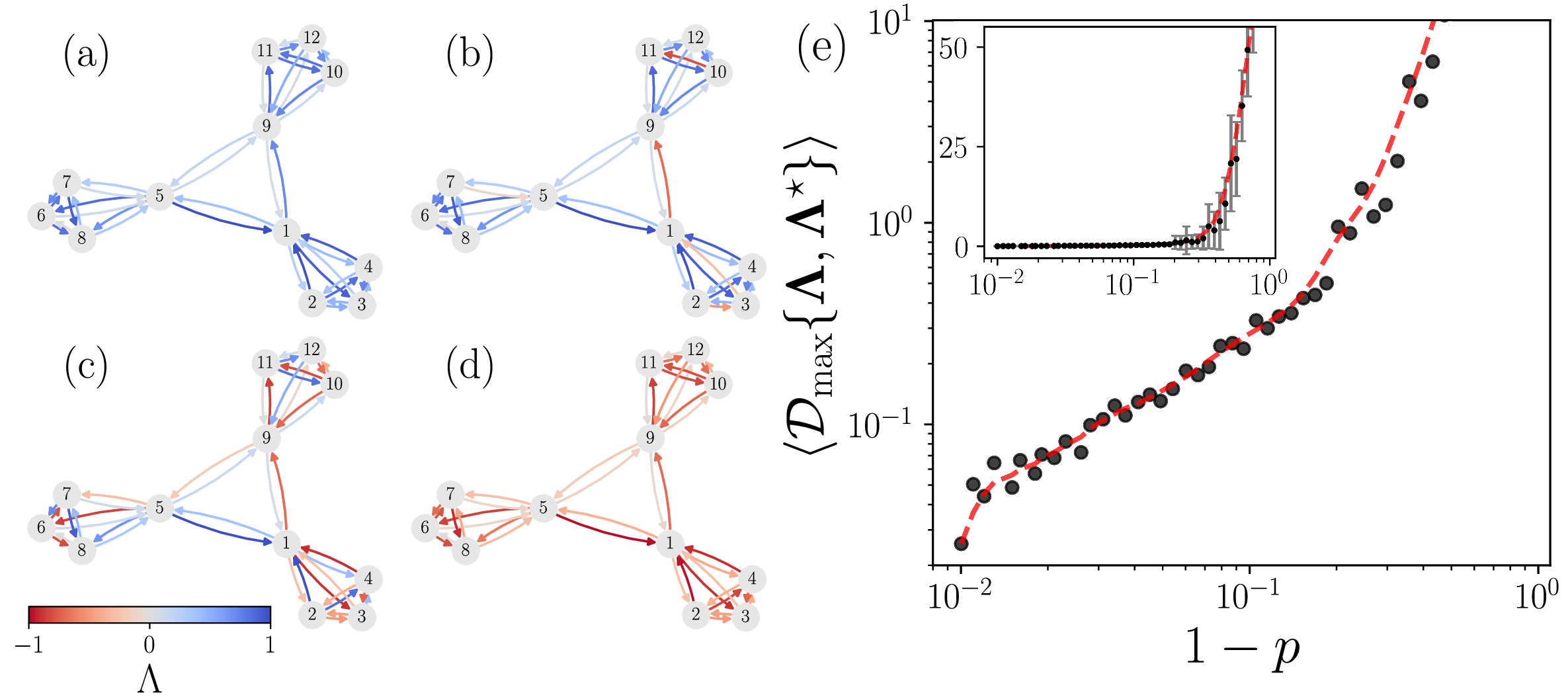}
	\vspace{1mm}
	\caption{Network with three cliques with a total of $N = 12$ nodes. The connection weights $\Lambda_{ij}$ are chosen randomly with a uniform distribution taking values between 0 and 1 as indicated in panel (a). In this structure, the sign is randomly changed, where $p$ is the probability of maintaining a positive sign and $1-p$ of changing to negative. The cases explored are: (a) $p=1.0$, (b) $p=0.8$, (c) $p=0.5$, (d) $p=0.0$. (e) Dissimilarity $\langle \mathcal{D}_{\mathrm{max}}(\bm{\Lambda}, \bm{\Lambda}^\star) \rangle$ as a function of the probability $1-p$ of sign change. The ensemble average $\langle \mathcal{D}_{\mathrm{max}}(\bm{\Lambda}, \bm{\Lambda}^\star) \rangle$ is calculated using 200 realizations for each value of $p$ and represented with a circle. The dashed line is a moving average that serves as a guide to observe the trend of the obtained values. The inset shows the results in semilogarithmic scale and the error bars represent the standard deviation of the data.}
	\label{Fig_5}
\end{figure}

Recent experimental advances have allowed the modeling of the human gut microbiome using twelve microbial populations whose dynamics follow a gLV model, with interactions described by a network  \cite{castellanos2023}. These experiments revealed that the interaction network exhibits community organization; that is, there exist groups of microorganisms with a higher density of intra-group interactions compared to inter-group ones. In addition, it is important to note that while population growth data is available from these experiments, such data alone is insufficient to uniquely determine all elements of the interaction matrix $\bm{\Lambda}$. To address this limitation, statistical methods based on Bayesian inference have been introduced to infer the matrix $\bm{\Lambda}$ \cite{Gibson2025}. These inference techniques have demonstrated that, beyond the network structure itself, the fraction of positive and negative entries in $\bm{\Lambda}$ plays a crucial role in determining the stability of the system \cite{castellanos2023}.
\\[2mm]
These findings motivate the study of the effects induced by the presence of both positive and negative interactions in a gLV system governed by Eqs.~(\ref{eq:GLV}) on a network with community structure. To explore this question using the dissimilarity measure between dynamical processes introduced in this work, we consider a network of $N = 12$ nodes arranged into three cliques of four nodes each. A clique is a subgraph in which every node is connected to all others \cite{Newman_Book2018}. These cliques are interconnected via a three-node ring, where each node is part of a different clique, thus forming a globally connected structure composed of three fully connected communities. In this idealized setup, the values of the non-zero entries $\Lambda_{ij}$ corresponding to the edges of the network are drawn independently from a uniform distribution in the interval $(0,1]$. Once a value $\Lambda_{ij}$ is assigned, its sign is determined stochastically: with probability $p$, the sign remains positive; with probability $1 - p$, the sign is flipped to negative.
\\[2mm]
In Fig. \ref{Fig_5} we analyze networks with cliques with $N = 12$ nodes. Figure \ref{Fig_5}(a) shows the network structure composed of three cliques. The color of each edge (as encoded in the color bar) represents the corresponding value of the matrix element $\Lambda_{ij}$ associated with that connection. If two nodes are not connected, then $\Lambda_{ij} = 0$. In this example, $p = 1.0$, so all edges have positive weights with random values in the interval $0 < \Lambda_{ij} \leq 1$. Panels \ref{Fig_5}(b)–(d) show the same network structure as in panel~(a), but with randomly assigned signs corresponding to $p = 0.8$, $p = 0.5$, and $p = 0.0$, respectively. In the configuration shown in Fig. \ref{Fig_5}(d), where $p = 0.0$, all edge weights are negative, but their absolute values match those in Fig. \ref{Fig_5}(a).
\\[2mm]
Once the network and the corresponding interaction strengths have been defined, it is important to understand the effect of the fraction of edges with negative signs on the resulting dynamics. To this end, we construct an interaction matrix $\bm{\Lambda}$ in which the nonzero entries, corresponding to the edges of the clique-based network [similar to the structure illustrated in Fig. \ref{Fig_5}(a)], are independently drawn from a uniform distribution in the interval $(0,1]$. As before, $\Lambda_{ij} = 0$ for pairs of nodes that are not connected.
\\[2mm]
The evolution of the values $\mathcal{N}_i(t)$ under the gLV equation with interaction matrix $\bm{\Lambda}$ is taken as the reference process. Based on this matrix, we define a second dynamical process with interaction matrix $\bm{\Lambda}^\star$, whose entries have the same absolute values as those in $\bm{\Lambda}$, but where each sign is independently kept positive with probability $p$ or flipped to negative with probability $1 - p$.
Using the same initial conditions as in the reference process, we solve numerically the modified dynamics to obtain $\mathcal{N}_i^\star(t)$. From this, we compute the dissimilarity $\mathcal{D}^{(\mathrm{LV})}(t)$ between the two processes, as defined in Eq.~(\ref{eq:Def_Dist_Red_genLV}). This allows us to determine the maximum dissimilarity $\mathcal{D}_{\mathrm{max}}\{\bm{\Lambda}, \bm{\Lambda}^\star\}$, defined in Eq.~(\ref{eq:dmax_red}). The value of $\mathcal{D}_{\mathrm{max}}\{\bm{\Lambda}, \bm{\Lambda}^\star\}$ can be computed repeatedly for the same probability $p$ controlling the sign perturbation in $\bm{\Lambda}^\star$. In each realization, we randomly generate the weights in the reference structure, the sign assignments in the perturbed matrix, and the initial values $\mathcal{N}_i(0)=\mathcal{N}_i^\star(0)$. The average over many such realizations, denoted by $\langle \mathcal{D}_{\mathrm{max}}\{\bm{\Lambda}, \bm{\Lambda}^\star\} \rangle$, thus provides a global statistical measure of the dissimilarity between the two processes when there is only a statistical knowledge about the values defining interaction matrices for the reference and the perturbed dynamics.
\\[2mm]
Figure \ref{Fig_5}(e) shows the ensemble average $\langle \mathcal{D}_{\mathrm{max}}\{\bm{\Lambda}, \bm{\Lambda}^\star\} \rangle$ as a function of the sign-flip probability $1 - p$. For each value of $p$, the average $\langle \mathcal{D}_{\mathrm{max}}\{\bm{\Lambda}, \bm{\Lambda}^\star\} \rangle$ was computed over $200$ realizations. In each realization, the interaction matrices $\bm{\Lambda}$ and $\bm{\Lambda}^\star$ for the reference and modified processes, respectively, were generated, and the gLV equations [Eq.~(\ref{eq:GLV})] were numerically integrated using a fourth-order Runge–Kutta algorithm. The resulting maximum dissimilarity $\mathcal{D}_{\mathrm{max}}\{\bm{\Lambda}, \bm{\Lambda}^\star\}$ was obtained for each realization. This procedure was repeated for a range of values of $p$. The results displayed in Fig. \ref{Fig_5}(e) illustrate the effect of sign perturbations in the interaction matrix. For small values of $1 - p$, the dissimilarity between the reference and modified processes remains low. Specifically, $\langle \mathcal{D}_{\mathrm{max}}\{\bm{\Lambda}, \bm{\Lambda}^\star\} \rangle$ increases gradually with $1 - p$, but remains small within the interval $0 < 1 - p \leq 0.2$, as highlighted in the inset. However, for $1 - p > 0.2$, the behavior of $\langle \mathcal{D}_{\mathrm{max}}\{\bm{\Lambda}, \bm{\Lambda}^\star\} \rangle$ changes drastically, exhibiting a rapid increase showing that the modified system behaves entirely differently from the stable reference system. For values $1 - p > 0.7$, the numerical integration of the modified system leads to divergence in the population abundances, indicating that a high fraction of negative entries in $\bm{\Lambda}^\star$ leads to an unstable system that diverges.
\subsection{Effect of modifications in a clique} 
\begin{figure}[b!]
	\centering
	\includegraphics*[width=0.95\textwidth]{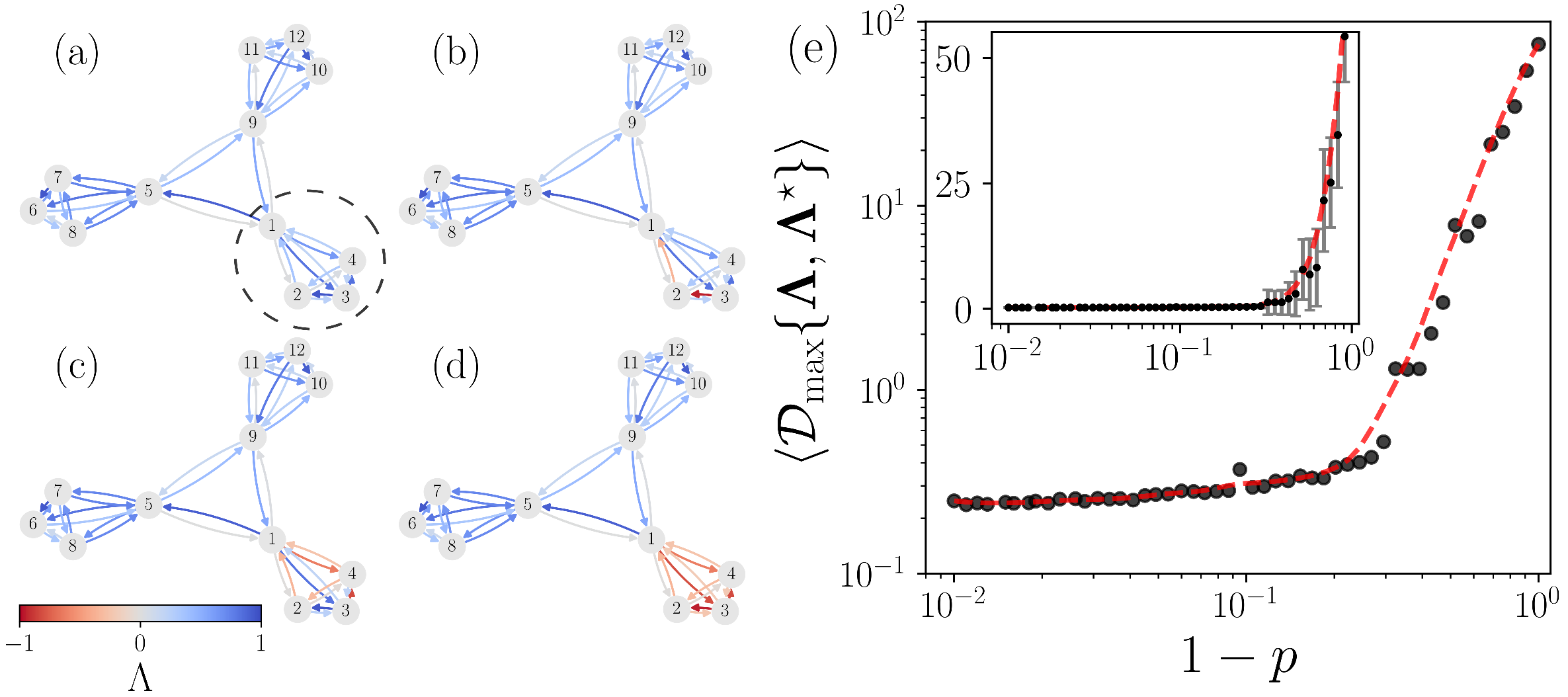}
	\vspace{1mm}
	\caption{Dissimilarity $\langle \mathcal{D}_{\mathrm{max}}\{\bm{\Lambda}, \bm{\Lambda}^\star\} \rangle$ as a function of the probability $1-p$ of sign-flip in the edges that are part of a clique. (a)-(d) present the structures analyzed. The sign is randomly changed in the links that are part of the clique formed by the nodes $i=1,2,3,4$, where $p$ is the probability of maintaining a positive sign and $1-p$ the probability of changing to negative. The cases explored are: (a) $p=1.0$, (b) $p=0.8$, (c) $p=0.5$, (d) $p=0.0$. (e) Ensemble average $\langle \mathcal{D}_{\mathrm{max}}\{\bm{\Lambda}, \bm{\Lambda}^\star\} \rangle$ calculated using 200 realizations for each value of $p$ and represented with circles. The dashed line is a moving average that serves as a guide to observe the trend of the obtained values. The inset shows the results on a semilogarithmic scale, and the error bars represent the standard deviation of the data.
	}
	\label{Fig_6}
\end{figure}
In the study of clique-based networks such as the one shown in Fig. \ref{Fig_5}(a), it is useful to understand the impact of localized changes within a single community (in this case, a clique). Motivated by this idea, in Fig. \ref{Fig_6} we analyze the effect of perturbations restricted to a single clique. The results are obtained by taking as reference a clique-based network similar to the one in Fig. \ref{Fig_6}(a), with a matrix $\bm{\Lambda}$ whose positive weights are randomly drawn from a uniform distribution in the interval $(0,1]$. For each realization of the reference process, we construct a modified process with an interaction matrix $\bm{\Lambda}^\star$, whose entries are identical to those in $\bm{\Lambda}$, except for those associated with the links within the clique formed by nodes $i = 1,2,3,4$. For these selected entries, the sign is flipped with probability $1 - p$, while the absolute values remain unchanged, as illustrated in Figs. \ref{Fig_6}(b)–(d). In both the reference and modified processes, the same randomly chosen initial conditions are used. This procedure is repeated for $200$ realizations, allowing us to compute the ensemble average $\langle \mathcal{D}_{\mathrm{max}}\{\bm{\Lambda}, \bm{\Lambda}^\star\} \rangle$. 
\\[2mm]
The results in Fig. \ref{Fig_6}(e) show $\langle \mathcal{D}_{\mathrm{max}}\{\bm{\Lambda}, \bm{\Lambda}^\star\} \rangle$ as a function of $1 - p$. A moving average (dashed line) is included to highlight the global trend, and an inset displays the data on a semi-logarithmic scale. The behavior observed in Fig. \ref{Fig_6}(e) for perturbations restricted to a single clique differs significantly from the global perturbations shown in Fig. \ref{Fig_5}(e). In particular, $\langle \mathcal{D}_{\mathrm{max}}\{\bm{\Lambda}, \bm{\Lambda}^\star\} \rangle$ remains nearly constant in the range $10^{-2} \leq 1 - p \leq 2 \times 10^{-1}$. For $1 - p > 2 \times 10^{-1}$, the dissimilarity increases rapidly, although it remains finite even as $1 - p \to 1$. This indicates that a complete sign inversion within a single clique does not destabilize the global dynamics of the system, and the values $\mathcal{N}^\star_i(t)$ remain stable. Here, it is important to emphasize that, in each realization, the full interaction matrix $\bm{\Lambda}$ is generated randomly. In particular, all nonzero weights, including those corresponding to the links connecting the altered clique to the other communities, are independently drawn from a uniform distribution in the interval $(0,1]$. Therefore, across realizations, the coupling strengths between the modified clique and the rest of the system span a broad range of possible values. As a consequence, the ensemble average $\langle \mathcal{D}_{\mathrm{max}}\{\bm{\Lambda}, \bm{\Lambda}^\star\} \rangle$ reported in Fig.~\ref{Fig_6}(e) incorporates variability not only in the internal clique interactions, but also in the inter-community connections. The observed robustness of the global dynamics under complete sign inversion within a single clique is therefore a statistical result, averaged over many realizations with heterogeneous coupling strengths to the rest of the network.
\subsection{Changes in the differential equations describing the modified system} 

\begin{figure}[b!]
	\centering
	\includegraphics*[width=0.9\textwidth]{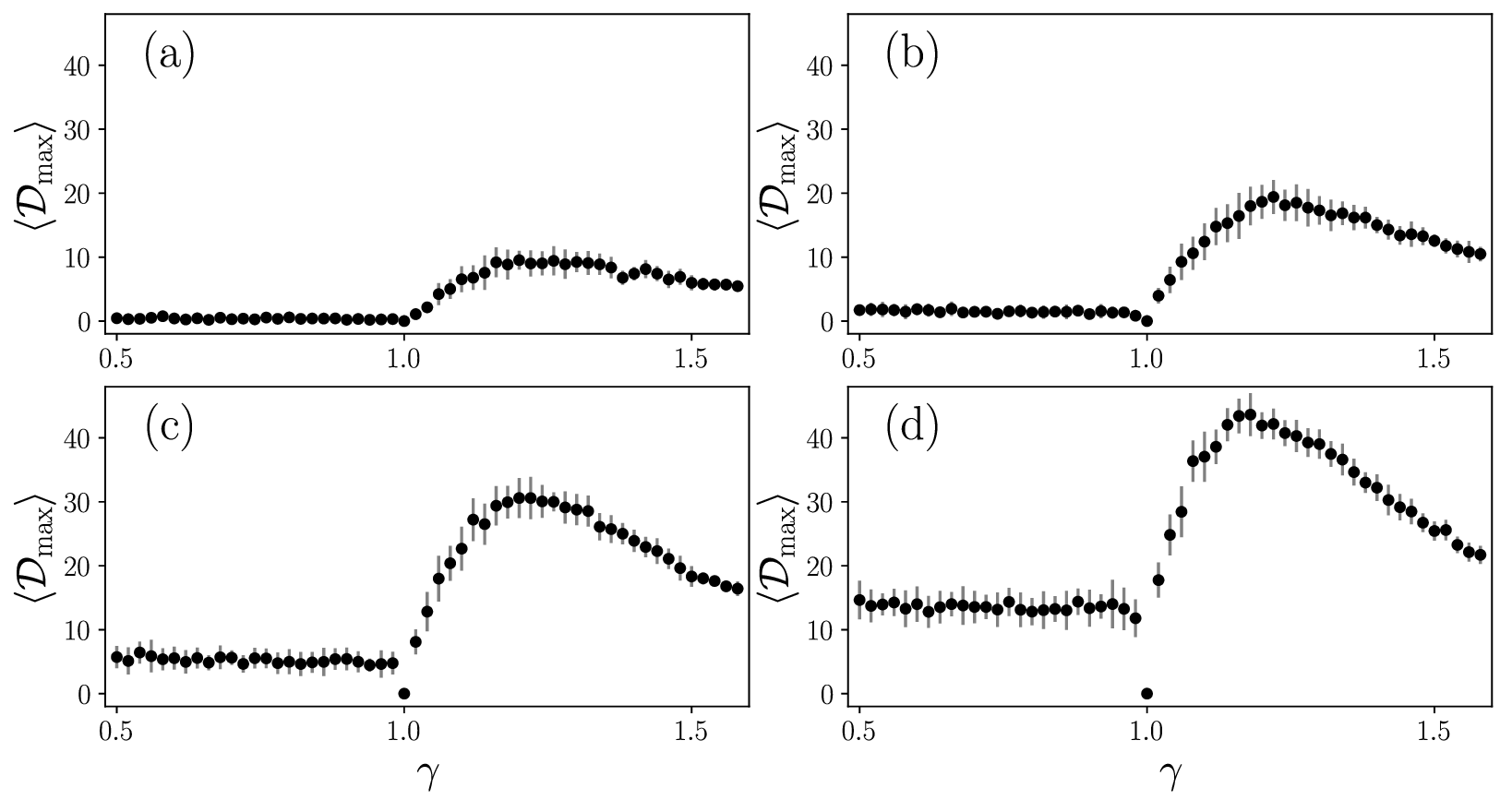}
	\vspace{-2mm}
	\caption{Dissimilarity $\langle \mathcal{D}_{\max} \rangle$ as a function of $\gamma$ for the modified dynamics governed by Eq.~(\ref{eq:GLV_gamma}) in random weighted networks with three cliques and size $N = 12$. The ensemble average $\langle \mathcal{D}_{\max} \rangle$ is calculated using 2000 realizations for each value of $\gamma$ and represented with circles. The error bars represent the standard deviation of the data. We consider random networks generated with different values of the probability of maintaining a positive sign with probability $p$ and changing to negative with probability $1-p$: (a) $p=0.8$, (b) $p=0.7$, (c) $p=0.6$, (b) $p=0.5$ (see the main text for details).}
	\label{Fig_7}
\end{figure}
All the cases analyzed so far have explored the effect of introducing changes in the interaction matrix $\bm{\Lambda}$ in order to define a modified system governed by $\bm{\Lambda}^\star$. In both the reference and modified processes, the temporal evolution is determined by the gLV equations defined in Eq.~(\ref{eq:GLV}). However, the formalism introduced is more general and can be applied to scenarios where the reference and modified systems evolve according to different types of differential equations. In this section, we consider a case in which both the reference and modified processes share the same interaction matrix $\bm{\Lambda}$. Starting from identical initial conditions, the reference process for $\mathcal{N}_i(t)$ evolves according to Eq.~(\ref{eq:GLV}), while the modified process for $\mathcal{N}^\star_i(t)$ follows the dynamics
\begin{equation}
	\label{eq:GLV_gamma}
	\dfrac{d \mathcal{N}^\star_i}{dt} = \mathcal{N}^\star_i \left(1 - \mathcal{N}^\star_i - \sum_{\substack{j=1\\ j\neq i}}^{N} \Lambda_{ij}\,(\mathcal{N}^\star_j)^\gamma \right),
\end{equation}
where $\gamma > 0$ is a real parameter that modifies the nonlinearity of the interactions between nodes. The case $\gamma = 1$ recovers the gLV equation used in the reference process.
\\[2mm]
Figure \ref{Fig_7} compares two processes for different values of $\gamma$ in random weighted networks with three cliques and size $N = 12$. For the case presented in Fig. \ref{Fig_7}(a), the interaction matrix $\bm{\Lambda}$ is randomly generated with $p = 0.8$, as illustrated in Fig. \ref{Fig_5}(b). The initial conditions for both processes are randomly generated but kept identical between $\mathcal{N}_i(0)$ and $\mathcal{N}^\star_i(0)$. For each realization of $\bm{\Lambda}$ and the initial conditions, we compute the dissimilarity $\mathcal{D}_{\mathrm{max}}$ by identifying the maximum of $\mathcal{D}^{(\mathrm{LV})}(t)$ defined in Eq.~(\ref{eq:Def_Dist_Red_genLV}). For each fixed value of $\gamma$, we consider $2000$ realizations, each allowing to obtain a value of $\mathcal{D}_{\mathrm{max}}$, from which we compute the ensemble average $\langle \mathcal{D}_{\mathrm{max}} \rangle$. The results are presented in Fig. \ref{Fig_7}(a) as a function of $\gamma$ over the interval $0.5 \leq \gamma <1.6$, where the error bars depict the standard deviation of the values. In Fig. \ref{Fig_7}(b)-(d), we present the same analysis for $p=0.7$, $p=0.6$, $p=0.5$, respectively.
\\[2mm]
The results for $\langle \mathcal{D}_{\mathrm{max}} \rangle$ in Fig. \ref{Fig_7} reveal a clear asymmetry in the effect of $\gamma$ around the reference value $\gamma = 1$, for which $\langle \mathcal{D}_{\mathrm{max}} \rangle = 0$ by construction. For the range of values explored, the dissimilarities $\langle \mathcal{D}_{\mathrm{max}} \rangle$ remain small for $\gamma < 1$, in contrast to the significantly larger values observed for $\gamma > 1$. The obtained results further highlight the importance of measures that enable the comparison of two dynamical processes. In the analyzed case, for example, $\langle \mathcal{D}_{\max} \rangle$ reveals an asymmetry with respect to $\gamma = 1$, which depends on both the dynamical processes and the weights defining the interactions.
\section{Discussion}
In this work, we introduced and applied a framework to quantify dissimilarities between generalized Lotka–Volterra dynamical processes at different levels of complexity, ranging from classical predator–prey systems to multispecies systems interacting through networks. By defining measures that capture both transient dynamics and stationary limits, we showed that changes in interaction parameters, network weights, or topology can lead to significant differences in the resulting dynamics. The proposed measures condense complex dynamical processes into different metrics, providing a systematic way to compare processes that differ in their interaction structure.
\\[2mm]
Our results highlight three main scenarios in which dynamical dissimilarities emerge. First, in two-species systems, the oscillatory dynamics are highly sensitive to both the interaction strength and the initial population sizes. Second, in directed three-node networks, we find that structural differences that appear indistinguishable at the level of adjacency matrices can nevertheless generate markedly different dynamical behaviors. This result emphasizes that structural metrics alone may be insufficient to characterize the similarity between interacting systems, and that dynamical comparisons are essential. Third, in networks with community structure, the fraction of negative interactions plays a decisive role in shaping the global dynamics. While small perturbations in the interaction signs preserve stability, beyond a critical threshold the dynamics change abruptly and may become unstable. Moreover, perturbations localized within a single clique produce different global responses than perturbations distributed across the entire network, highlighting the importance of the spatial organization of interactions.
\\[2mm]
The results also show that the proposed framework is general and enables comparisons between dynamical processes governed by different nonlinear interaction terms. This flexibility makes the approach particularly relevant for ecological and microbiome studies, where interaction matrices are often inferred from experimental data and therefore subject to uncertainty, or where alternative functional forms may be considered to describe species interactions. Importantly, the proposed dissimilarity measures provide a quantitative way to assess how sensitive the resulting dynamics are to such modifications. By tracking the divergence between a reference process and perturbed dynamics, the framework makes it possible to identify parameter regimes or structural changes that produce large dynamical deviations or lead to divergent population trajectories. In this way, the measures can reveal thresholds beyond which small perturbations in interaction strengths, signs, or nonlinearities trigger substantial changes in system behavior, thereby helping to identify conditions under which ecosystems or microbial communities may lose stability. We emphasize that the simplifying assumptions adopted in specific examples, such as identical intrinsic growth rates, were introduced for clarity and do not restrict the applicability of the framework, which can be readily extended to more general settings with heterogeneous species-specific parameters.
\\[2mm]
This study introduces a methodological framework for the comparative analysis of complex systems modeled by nonlinear coupled differential equations. In contrast to approaches that rely exclusively on structural comparisons of interaction matrices, the proposed method focuses on the resulting dynamics, allowing the identification of divergences that emerge from nonlinear interactions, transient behaviors, or alternative dynamical formulations. A limitation of the approach is that it requires numerical integration of the governing equations, which can be computationally more demanding than structural metrics computed directly from the interaction matrix, particularly when ensemble analyses over interaction matrices and initial conditions are considered. For this reason, the numerical experiments presented in this work focus on small and intermediate-size systems, where the mechanisms generating the observed dynamical differences can be directly interpreted. Although this choice facilitates a clear illustration of the proposed methodology, extending the framework to much larger networks remains an important direction for future research. In particular, applying the method to empirically derived interaction networks, such as ecological communities, microbial consortia, or neural systems, may require the development of approximate or reduced descriptions of the dynamics capable of preserving the relevant dynamical information while remaining computationally tractable. Despite these limitations, purely structural measures cannot fully capture how nonlinear dynamics shape the evolution of interacting populations or how small perturbations may propagate through the system over time. In this sense, the framework presented here represents a first numerical step toward systematically characterizing divergences between dynamical processes. By linking structural perturbations and model variations to their dynamical consequences, the proposed measures provide a foundation for future approaches aimed at identifying regimes of stability, resilience, and instability in complex interacting systems.

\section*{Author Contributions}
M.C.D.M. and A.P.R. conceived the basic research idea. N.A.M. and A.P.R carried out the analysis and drafted the paper. All authors revised the paper.

\section*{Competing interests}
The authors declare no competing interests.

\section*{Funding}
This research did not receive any specific grant from funding agencies in the public, commercial, or not for profit sectors.

\section*{Data availability}
All data, models, and code used to support the findings of this study are available from the corresponding author upon reasonable request.


\begin{thebibliography}{10}
\expandafter\ifx\csname url\endcsname\relax
  \def\url#1{\texttt{#1}}\fi
\expandafter\ifx\csname urlprefix\endcsname\relax\def\urlprefix{URL }\fi
\expandafter\ifx\csname doiprefix\endcsname\relax\def\doiprefix{DOI }\fi
\providecommand{\bibinfo}[2]{#2}
\providecommand{\eprint}[2][]{\url{#2}}

\bibitem{Riva2023EcologicalComplexity}
\bibinfo{author}{Riva, F.} \emph{et~al.}
\newblock \bibinfo{journal}{\bibinfo{title}{Toward a cohesive understanding of
  ecological complexity}}.
\newblock {\emph{\JournalTitle{Sci. Adv.}}} \textbf{\bibinfo{volume}{9}},
  \bibinfo{pages}{eabq4207} (\bibinfo{year}{2023}).
\newblock \doiprefix 10.1126/sciadv.abq4207.

\bibitem{SoleLevin2022EcologicalComplexity}
\bibinfo{author}{Sol{\'e}, R.} \& \bibinfo{author}{Levin, S.}
\newblock \bibinfo{journal}{\bibinfo{title}{Ecological complexity and the
  biosphere: the next 30 years}}.
\newblock {\emph{\JournalTitle{Philos. Trans. R. Soc. B Biol. Sci.}}}
  \textbf{\bibinfo{volume}{377}}, \bibinfo{pages}{20210376}
  (\bibinfo{year}{2022}).
\newblock \doiprefix 10.1098/rstb.2021.0376.

\bibitem{Fortin2021NetworkEcology}
\bibinfo{author}{Fortin, M.-J.}, \bibinfo{author}{Dale, M.~R.} \&
  \bibinfo{author}{Brimacombe, C.}
\newblock \bibinfo{journal}{\bibinfo{title}{Network ecology in dynamic
  landscapes}}.
\newblock {\emph{\JournalTitle{Proc. R. Soc. B: Biol. Sci.}}}
  \textbf{\bibinfo{volume}{288}}, \bibinfo{pages}{20201889}
  (\bibinfo{year}{2021}).
\newblock \doiprefix 10.1098/rspb.2020.1889.

\bibitem{Dormann2017PatternsEcologicalNetworks}
\bibinfo{author}{Dormann, C.~F.}, \bibinfo{author}{Fründ, J.} \&
  \bibinfo{author}{Schaefer, H.~M.}
\newblock \bibinfo{journal}{\bibinfo{title}{Identifying causes of patterns in
  ecological networks: Opportunities and limitations}}.
\newblock {\emph{\JournalTitle{Annu. Rev. Ecol. Evol. Syst.}}}
  \textbf{\bibinfo{volume}{48}}, \bibinfo{pages}{559--584}
  (\bibinfo{year}{2017}).
\newblock \doiprefix 10.1146/annurev-ecolsys-110316-022928.

\bibitem{Pilosof2017MultilayerEcologicalNetworks}
\bibinfo{author}{Pilosof, S.}, \bibinfo{author}{Porter, M.~A.},
  \bibinfo{author}{Pascual, M.} \& \bibinfo{author}{Kéfi, S.}
\newblock \bibinfo{journal}{\bibinfo{title}{The multilayer nature of ecological
  networks}}.
\newblock {\emph{\JournalTitle{Nat. Ecol. Evol.}}}
  \textbf{\bibinfo{volume}{1}}, \bibinfo{pages}{0101} (\bibinfo{year}{2017}).
\newblock \doiprefix 10.1038/s41559-017-0101.

\bibitem{PhysRevE.111.034408}
\bibinfo{author}{Kessler, D.~A.} \& \bibinfo{author}{Shnerb, N.~M.}
\newblock \bibinfo{journal}{\bibinfo{title}{Interaction network structures in
  competitive ecosystems}}.
\newblock {\emph{\JournalTitle{Phys. Rev. E}}} \textbf{\bibinfo{volume}{111}},
  \bibinfo{pages}{034408} (\bibinfo{year}{2025}).
\newblock \doiprefix 10.1103/PhysRevE.111.034408.

\bibitem{PhysRevE.111.014318}
\bibinfo{author}{Poley, L.}, \bibinfo{author}{Galla, T.} \&
  \bibinfo{author}{Baron, J.~W.}
\newblock \bibinfo{journal}{\bibinfo{title}{{Interaction networks in persistent
  Lotka-Volterra communities}}}.
\newblock {\emph{\JournalTitle{Phys. Rev. E}}} \textbf{\bibinfo{volume}{111}},
  \bibinfo{pages}{014318} (\bibinfo{year}{2025}).
\newblock \doiprefix 10.1103/PhysRevE.111.014318.

\bibitem{Lastad2022}
\bibinfo{author}{L{\aa}stad, S.~B.} \& \bibinfo{author}{Haerter, J.~O.}
\newblock \bibinfo{journal}{\bibinfo{title}{The geometry of evolved community
  matrix spectra}}.
\newblock {\emph{\JournalTitle{Sci. Rep.}}} \textbf{\bibinfo{volume}{12}},
  \bibinfo{pages}{14668} (\bibinfo{year}{2022}).
\newblock \doiprefix 10.1038/s41598-022-17379-6.

\bibitem{10.1111/2041-210X.70032}
\bibinfo{author}{Pascal, L.~V.} \emph{et~al.}
\newblock \bibinfo{journal}{\bibinfo{title}{{EEMtoolbox: A user-friendly R
  package for flexible ensemble ecosystem modelling}}}.
\newblock {\emph{\JournalTitle{Methods Ecol. Evol.}}}
  \textbf{\bibinfo{volume}{16}}, \bibinfo{pages}{921--929}
  (\bibinfo{year}{2025}).
\newblock \doiprefix 10.1111/2041-210X.70032.

\bibitem{Stein2013}
\bibinfo{author}{Stein, R.~R.} \emph{et~al.}
\newblock \bibinfo{journal}{\bibinfo{title}{Ecological modeling from
  time-series inference: Insight into dynamics and stability of intestinal
  microbiota}}.
\newblock {\emph{\JournalTitle{PLoS Comput. Biol.}}}
  \textbf{\bibinfo{volume}{9}}, \bibinfo{pages}{e1003388}
  (\bibinfo{year}{2013}).
\newblock \doiprefix 10.1371/journal.pcbi.1003388.

\bibitem{Bucci2016}
\bibinfo{author}{Bucci, V.} \emph{et~al.}
\newblock \bibinfo{journal}{\bibinfo{title}{Mdsine: Microbial dynamical systems
  inference engine for microbiome time-series}}.
\newblock {\emph{\JournalTitle{Genome Biol.}}} \textbf{\bibinfo{volume}{17}},
  \bibinfo{pages}{121} (\bibinfo{year}{2016}).
\newblock \doiprefix 10.1186/s13059-016-0980-6.

\bibitem{Venturelli2018}
\bibinfo{author}{Venturelli, O.~S.} \emph{et~al.}
\newblock \bibinfo{journal}{\bibinfo{title}{Deciphering microbial interactions
  in synthetic human gut communities}}.
\newblock {\emph{\JournalTitle{Mol. Syst. Biol.}}}
  \textbf{\bibinfo{volume}{14}}, \bibinfo{pages}{e8157} (\bibinfo{year}{2018}).
\newblock \doiprefix 10.15252/msb.20178157.

\bibitem{10.1016/j.fm.2025.104767}
\bibinfo{author}{Liu, C.}, \bibinfo{author}{Yi, F.}, \bibinfo{author}{Niu, C.}
  \& \bibinfo{author}{Li, Q.}
\newblock \bibinfo{journal}{\bibinfo{title}{Unravelling microbial interactions
  in a synthetic broad bean paste microbial community}}.
\newblock {\emph{\JournalTitle{Food Microbiol.}}}
  \textbf{\bibinfo{volume}{130}} (\bibinfo{year}{2025}).
\newblock \doiprefix 10.1016/j.fm.2025.104767.

\bibitem{Melvan_2025}
\bibinfo{author}{Melvan, E.}, \bibinfo{author}{Allen, A.~P.},
  \bibinfo{author}{Vuckovic, T.}, \bibinfo{author}{Soljic, I.} \&
  \bibinfo{author}{Starčević, A.}
\newblock \bibinfo{journal}{\bibinfo{title}{Predicting gut microbiota dynamics
  in obese individuals from cross-sectional data}}.
\newblock {\emph{\JournalTitle{Front. Cell. Infect. Microbiol.}}}
  \textbf{\bibinfo{volume}{15}}, \bibinfo{pages}{1485791}
  (\bibinfo{year}{2025}).
\newblock \doiprefix 10.3389/fcimb.2025.1485791.

\bibitem{Gibson2025}
\bibinfo{author}{Gibson, T.~E.} \emph{et~al.}
\newblock \bibinfo{journal}{\bibinfo{title}{{Learning ecosystem-scale dynamics
  from microbiome data with MDSINE2}}}.
\newblock {\emph{\JournalTitle{Nat. Microbiol.}}}
  \textbf{\bibinfo{volume}{10}}, \bibinfo{pages}{2550--2564}
  (\bibinfo{year}{2025}).
\newblock \doiprefix 10.1038/s41564-025-02112-6.

\bibitem{castellanos2023}
\bibinfo{author}{Castellanos, A.}
\newblock \emph{\bibinfo{title}{Towards an Ecological and Functional Framework
  for Modeling the Structure and Dynamics of the Human Gut Microbiome}}.
\newblock \bibinfo{type}{Master's thesis}, \bibinfo{school}{Universidad de los
  Andes}, \bibinfo{address}{Bogotá, Colombia} (\bibinfo{year}{2023}).

\bibitem{Lagzi2015}
\bibinfo{author}{Lagzi, F.} \& \bibinfo{author}{Rotter, S.}
\newblock \bibinfo{journal}{\bibinfo{title}{Dynamics of competition between
  subnetworks of spiking neuronal networks in the balanced state}}.
\newblock {\emph{\JournalTitle{PLoS One}}} \textbf{\bibinfo{volume}{10}},
  \bibinfo{pages}{e0138947} (\bibinfo{year}{2015}).
\newblock \doiprefix 10.1371/journal.pone.0138947.

\bibitem{MeyerOrtmanns2023}
\bibinfo{author}{Meyer-Ortmanns, H.}
\newblock \bibinfo{journal}{\bibinfo{title}{Heteroclinic networks for brain
  dynamics}}.
\newblock {\emph{\JournalTitle{Front. Netw. Physiol.}}}
  \textbf{\bibinfo{volume}{3}}, \bibinfo{pages}{1276401}
  (\bibinfo{year}{2023}).
\newblock \doiprefix 10.3389/fnetp.2023.1276401.

\bibitem{Rabinovich2024}
\bibinfo{author}{Rabinovich, M.}, \bibinfo{author}{Bick, C.} \&
  \bibinfo{author}{Varona, P.}
\newblock \bibinfo{journal}{\bibinfo{title}{Beyond neurons and spikes: cognon,
  the hierarchical dynamical unit of thought}}.
\newblock {\emph{\JournalTitle{Cogn. Neurodyn.}}}
  \textbf{\bibinfo{volume}{18}}, \bibinfo{pages}{3327--3335}
  (\bibinfo{year}{2024}).
\newblock \doiprefix 10.1007/s11571-023-09987-3.

\bibitem{10_1098_rspa_2024_0290}
\bibinfo{author}{Mooij, M.~N.}, \bibinfo{author}{Baudena, M.},
  \bibinfo{author}{von~der Heydt, A.~S.} \& \bibinfo{author}{Kryven, I.}
\newblock \bibinfo{journal}{\bibinfo{title}{Stable coexistence in indefinitely
  large systems of competing species}}.
\newblock {\emph{\JournalTitle{Proc. R. Soc. A: Math. Phys. Eng. Sci.}}}
  \textbf{\bibinfo{volume}{480}}, \bibinfo{pages}{20240290}
  (\bibinfo{year}{2024}).
\newblock \doiprefix 10.1098/rspa.2024.0290.

\bibitem{10_1016_j_physrep_2024_08_001}
\bibinfo{author}{Chen, C.}, \bibinfo{author}{Wang, X.-W.} \&
  \bibinfo{author}{Liu, Y.-Y.}
\newblock \bibinfo{journal}{\bibinfo{title}{Stability of ecological systems: A
  theoretical review}}.
\newblock {\emph{\JournalTitle{Phys. Rep.}}} \textbf{\bibinfo{volume}{1088}},
  \bibinfo{pages}{1--41} (\bibinfo{year}{2024}).
\newblock \doiprefix 10.1016/j.physrep.2024.08.001.

\bibitem{PhysRevE.110.044309}
\bibinfo{author}{Eraso-Hernandez, L.~K.} \& \bibinfo{author}{Riascos, A.~P.}
\newblock \bibinfo{journal}{\bibinfo{title}{Antifragility of stochastic
  transport on networks with damage}}.
\newblock {\emph{\JournalTitle{Phys. Rev. E}}} \textbf{\bibinfo{volume}{110}},
  \bibinfo{pages}{044309} (\bibinfo{year}{2024}).
\newblock \doiprefix 10.1103/PhysRevE.110.044309.

\bibitem{PoloGonzalez2025}
\bibinfo{author}{Polo-González, M.~A.}, \bibinfo{author}{Riascos, A.~P.} \&
  \bibinfo{author}{Eraso-Hernandez, L.~K.}
\newblock \bibinfo{journal}{\bibinfo{title}{Antifragility and response to
  damage in the synchronization of oscillators on networks}}.
\newblock {\emph{\JournalTitle{J. Phys. A: Math. Theor.}}}
  \textbf{\bibinfo{volume}{58}}, \bibinfo{pages}{225002}
  (\bibinfo{year}{2025}).
\newblock \doiprefix 10.1088/1751-8121/add974.

\bibitem{VespiBook}
\bibinfo{author}{Barrat, A.}, \bibinfo{author}{Barth\'elemy, M.} \&
  \bibinfo{author}{Vespignani, A.}
\newblock \emph{\bibinfo{title}{Dynamical Processes on Complex Networks}}
  (\bibinfo{publisher}{Cambridge University Press},
  \bibinfo{address}{Cambridge}, \bibinfo{year}{2008}).

\bibitem{Barabasi_2016}
\bibinfo{author}{Barabási, A.-L.}
\newblock \emph{\bibinfo{title}{Network Science}}
  (\bibinfo{publisher}{Cambridge University Press}, \bibinfo{year}{2016}).

\bibitem{Newman_Book2018}
\bibinfo{author}{Newman, M.}
\newblock \emph{\bibinfo{title}{Networks: An Introduction}}
  (\bibinfo{publisher}{Oxford University Press, Inc.}, \bibinfo{address}{USA},
  \bibinfo{year}{2018}).

\bibitem{Hamming1950}
\bibinfo{author}{Hamming, R.~W.}
\newblock \bibinfo{journal}{\bibinfo{title}{Error detecting and error
  correcting codes}}.
\newblock {\emph{\JournalTitle{Bell Syst. Tech. J.}}}
  \textbf{\bibinfo{volume}{29}}, \bibinfo{pages}{147--160}
  (\bibinfo{year}{1950}).
\newblock \doiprefix 10.1002/j.1538-7305.1950.tb00463.x.

\bibitem{Jaccard1901}
\bibinfo{author}{Jaccard, P.}
\newblock \bibinfo{journal}{\bibinfo{title}{Etude de la distribution florale
  dans une portion des alpes et du jura}}.
\newblock {\emph{\JournalTitle{Bull. de la Soc. Vaud. des Sci. Nat.}}}
  \textbf{\bibinfo{volume}{37}}, \bibinfo{pages}{547--579}
  (\bibinfo{year}{1901}).
\newblock \doiprefix 10.5169/seals-266450.

\bibitem{LEVANDOWSKY1971}
\bibinfo{author}{Levandowsky, M.} \& \bibinfo{author}{Winter, D.}
\newblock \bibinfo{journal}{\bibinfo{title}{Distance between sets}}.
\newblock {\emph{\JournalTitle{Nature}}} \textbf{\bibinfo{volume}{234}},
  \bibinfo{pages}{34--35} (\bibinfo{year}{1971}).
\newblock \doiprefix 10.1038/234034a0.

\bibitem{Bagrow2019}
\bibinfo{author}{Bagrow, J.~P.} \& \bibinfo{author}{Bollt, E.~M.}
\newblock \bibinfo{journal}{\bibinfo{title}{An information-theoretic,
  all-scales approach to comparing networks}}.
\newblock {\emph{\JournalTitle{Appl. Netw. Sci.}}}
  \textbf{\bibinfo{volume}{4}}, \bibinfo{pages}{45} (\bibinfo{year}{2019}).
\newblock \doiprefix 10.1007/s41109-019-0156-x.

\bibitem{Donnat_2018}
\bibinfo{author}{Donnat, C.} \& \bibinfo{author}{Holmes, S.}
\newblock \bibinfo{journal}{\bibinfo{title}{{Tracking network dynamics: A
  survey using graph distances}}}.
\newblock {\emph{\JournalTitle{Ann. Appl. Stat.}}}
  \textbf{\bibinfo{volume}{12}}, \bibinfo{pages}{971 -- 1012}
  (\bibinfo{year}{2018}).
\newblock \doiprefix 10.1214/18-AOAS1176.

\bibitem{Jurman2015}
\bibinfo{author}{Jurman, G.}, \bibinfo{author}{Visintainer, R.},
  \bibinfo{author}{Filosi, M.}, \bibinfo{author}{Riccadonna, S.} \&
  \bibinfo{author}{Furlanello, C.}
\newblock \bibinfo{title}{{The HIM glocal metric and kernel for network
  comparison and classification}}.
\newblock In \emph{\bibinfo{booktitle}{2015 IEEE International Conference on
  Data Science and Advanced Analytics (DSAA)}}, \bibinfo{pages}{1--10}
  (\bibinfo{publisher}{IEEE}, \bibinfo{year}{2015}).
\newblock \doiprefix 10.1109/DSAA.2015.7344816.

\bibitem{Hammond2013}
\bibinfo{author}{Hammond, D.~K.}, \bibinfo{author}{Gur, Y.} \&
  \bibinfo{author}{Johnson, C.~R.}
\newblock \bibinfo{title}{Graph diffusion distance: A difference measure for
  weighted graphs based on the graph laplacian exponential kernel}.
\newblock In \emph{\bibinfo{booktitle}{2013 IEEE Global Conference on Signal
  and Information Processing}}, \bibinfo{pages}{419--422}
  (\bibinfo{year}{2013}).
\newblock \doiprefix 10.1109/GlobalSIP.2013.6736904.

\bibitem{Scott2021}
\bibinfo{author}{Scott, C.~B.} \& \bibinfo{author}{Mjolsness, E.}
\newblock \bibinfo{journal}{\bibinfo{title}{Graph diffusion distance:
  Properties and efficient computation}}.
\newblock {\emph{\JournalTitle{PLoS One}}} \textbf{\bibinfo{volume}{16}},
  \bibinfo{pages}{e0249624} (\bibinfo{year}{2021}).
\newblock \doiprefix 10.1371/journal.pone.0249624.

\bibitem{DomenicoPRE2020}
\bibinfo{author}{Ghavasieh, A.}, \bibinfo{author}{Nicolini, C.} \&
  \bibinfo{author}{De~Domenico, M.}
\newblock \bibinfo{journal}{\bibinfo{title}{Statistical physics of complex
  information dynamics}}.
\newblock {\emph{\JournalTitle{Phys. Rev. E}}} \textbf{\bibinfo{volume}{102}},
  \bibinfo{pages}{052304} (\bibinfo{year}{2020}).
\newblock \doiprefix 10.1103/PhysRevE.102.052304.

\bibitem{RiascosDiffusion2023}
\bibinfo{author}{Riascos, A.~P.} \& \bibinfo{author}{Padilla, F.~H.}
\newblock \bibinfo{journal}{\bibinfo{title}{A measure of dissimilarity between
  diffusive processes on networks}}.
\newblock {\emph{\JournalTitle{J. Phys. A: Math. Theor.}}}
  \textbf{\bibinfo{volume}{56}}, \bibinfo{pages}{145001}
  (\bibinfo{year}{2023}).
\newblock \doiprefix 10.1088/1751-8121/acc144.

\bibitem{RiascosKuramoto2023}
\bibinfo{author}{Riascos, A.~P.}
\newblock \bibinfo{journal}{\bibinfo{title}{Dissimilarity between
  synchronization processes on networks}}.
\newblock {\emph{\JournalTitle{Phys. Rev. E}}} \textbf{\bibinfo{volume}{109}},
  \bibinfo{pages}{044301} (\bibinfo{year}{2024}).
\newblock \doiprefix 10.1103/PhysRevE.109.044301.

\bibitem{Muscarella2020}
\bibinfo{author}{Muscarella, M.~E.} \& \bibinfo{author}{O'Dwyer, J.~P.}
\newblock \bibinfo{journal}{\bibinfo{title}{Species dynamics and interactions
  via metabolically informed consumer-resource models}}.
\newblock {\emph{\JournalTitle{Theor. Ecol.}}} \textbf{\bibinfo{volume}{13}},
  \bibinfo{pages}{503--518} (\bibinfo{year}{2020}).
\newblock \doiprefix 10.1007/s12080-020-00466-7.

\bibitem{Murray2002}
\bibinfo{author}{Murray, J.~D.}
\newblock \emph{\bibinfo{title}{Mathematical Biology I: An Introduction}},
  vol.~\bibinfo{volume}{17} of \emph{\bibinfo{series}{Interdisciplinary Applied
  Mathematics}} (\bibinfo{publisher}{Springer}, \bibinfo{address}{New York},
  \bibinfo{year}{2002}), \bibinfo{edition}{3} edn.

\bibitem{VOLTERRA1926}
\bibinfo{author}{Volterra, V. I. T.~O.}
\newblock \bibinfo{journal}{\bibinfo{title}{Fluctuations in the abundance of a
  species considered mathematically}}.
\newblock {\emph{\JournalTitle{Nature}}} \textbf{\bibinfo{volume}{118}},
  \bibinfo{pages}{558--560} (\bibinfo{year}{1926}).
\newblock \doiprefix 10.1038/118558a0.

\bibitem{lotka1925elements}
\bibinfo{author}{Lotka, A.}
\newblock \emph{\bibinfo{title}{Elements of Physical Biology}}
  (\bibinfo{publisher}{Williams \& Wilkins}, \bibinfo{year}{1925}).

\bibitem{Kot_2001}
\bibinfo{author}{Kot, M.}
\newblock \emph{\bibinfo{title}{Elements of Mathematical Ecology}}
  (\bibinfo{publisher}{Cambridge University Press}, \bibinfo{year}{2001}).

\bibitem{Liu2023}
\bibinfo{author}{Liu, X.}, \bibinfo{author}{Constable, G. W.~A.} \&
  \bibinfo{author}{Pitchford, J.~W.}
\newblock \bibinfo{journal}{\bibinfo{title}{{Feasibility and stability in large
  Lotka Volterra systems with interaction structure}}}.
\newblock {\emph{\JournalTitle{Phys. Rev. E}}} \textbf{\bibinfo{volume}{107}},
  \bibinfo{pages}{054301} (\bibinfo{year}{2023}).
\newblock \doiprefix 10.1103/PhysRevE.107.054301.

\bibitem{may1972}
\bibinfo{author}{May, R.}
\newblock \bibinfo{journal}{\bibinfo{title}{Will a large complex system be
  stable?}}
\newblock {\emph{\JournalTitle{Nature}}} \textbf{\bibinfo{volume}{238}},
  \bibinfo{pages}{413--414} (\bibinfo{year}{1972}).

\bibitem{Akjouj2024}
\bibinfo{author}{Akjouj, I.} \emph{et~al.}
\newblock \bibinfo{journal}{\bibinfo{title}{Complex systems in ecology: a
  guided tour with large {L}otka–{V}olterra models and random matrices}}.
\newblock {\emph{\JournalTitle{Proc. R. Soc. A: Math. Phys. Eng. Sci.}}}
  \textbf{\bibinfo{volume}{480}}, \bibinfo{pages}{20230284}
  (\bibinfo{year}{2024}).
\newblock \doiprefix 10.1098/rspa.2023.0284.

\bibitem{Allesina2012}
\bibinfo{author}{Allesina, S.} \& \bibinfo{author}{Tang, S.}
\newblock \bibinfo{journal}{\bibinfo{title}{Stability criteria for complex
  ecosystems}}.
\newblock {\emph{\JournalTitle{Nature}}} \textbf{\bibinfo{volume}{483}},
  \bibinfo{pages}{205--208} (\bibinfo{year}{2012}).
\newblock \doiprefix 10.1038/nature10832.

\end{thebibliography}

\end{document}